# Search for Possible Solar Influences in Ra-226 Decays

Daniel D. Stancil[1], Sümeyra Balci Yegen[2,†], David A. Dickey[3], and Chris R. Gould[2]
[1]Department of Electrical and Computer Engineering
[2]Department of Physics
[3]Department of Statistics
North Carolina State University
Raleigh, NC 27695

**Abstract**

Measurements of Ra-226 activity from eight HPGe gamma ray detectors at the NC State University PULSTAR Reactor were analyzed for evidence of periodic variations, with particular attention to annual variations. All measurements were made using the same reference source, and data sets were of varying length taken over the time period from September 1996 through August 2014. Clear evidence of annual variations was observed in data from four of the detectors. Short time periodograms from the data sets suggest temporal variability of both the amplitude and frequency of these variations. The annual variations in two of the data sets show peak values near the first of February, while surprisingly, the annual variations in the other two are roughly out of phase with the first two. Three of the four detectors exhibited annual variations over approximately the same time period. A joint statistic constructed by combining spectra from these three shows peaks approximating the frequencies of solar r-mode oscillations with $\nu_R = 11.74 \, \text{cpy}$, $m = 1$, and $l = 3, 5, 6$. The fact that similar variations were not present in all detectors covering similar time periods rules out variations in activity as the cause, and points to differing sensitivities to unspecified environmental parameters instead. In addition to seasonal variations, the modulation of environmental parameters by solar processes remains a possible explanation of periodogram features, but without requiring new physics.

## 1. Introduction

During the 1980s and 1990s, examination of radioactivity data taken over multiple year periods to determine the lifetime of several long-lived isotopes resulted in the observation of apparent annual variations in the measured activity [1-3]. Further, these variations tended to be either in phase, or in anti-phase with the seasons.

Independently, Falkenberg [4] performed an experiment to specifically look for annual variations in the decay rate of tritium that correlated with the distance between the earth and the sun, based on the hypothesis that radioactive decay might be caused by neutrinos. An annual variation that was roughly in phase with the earth-sun distance was indeed observed, in spite of measures taken to ensure shielding of environmental variables such as temperature, light, and other forms of electromagnetic radiation.

---

[†] Current address: Department of Physics, Çukurova University, Adana, Turkey.

During the course of designing an experiment to statistically test the randomness of radioactive decay, Fischbach and coworkers ran across the results obtained by Alburger et al at the Brookhaven National Laboratory (BNL) [1], and by Siegert et al at the Physikalisch-Technische Bundesanstalt (PTB) in Germany [3]. In a series of papers beginning in 2009, Fischbach, Jenkins, Sturrock and collaborators presented analyses of the BNL and PTB data and argued that it is unlikely that local environmental effects are responsible for the annual variations observed, thus favoring the hypothesis that some type of radiation from the sun was responsible owing to the change in earth-sun distance [5-12]. Evidence for annual variations has also been observed in nuclear decay data from the Lomonosov Moscow State University [13-15], the Geological Survey of Israel (GSI) Laboratory in Jerusalem [16], [17], the Ohio State University Research Reactor [18], and the Baksan Neutrino Observatory of the Institute for Nuclear Researches [19]. An annual variation was also observed at the Gran Sasso Laboratory (LNGS), but was attributed to the cosmic ray background [20].

In contrast, several groups have looked for evidence of an annual variation in decay rates and obtained a negative result [20-25]. It should be noted, however, that the data set that was the basis for the negative conclusion by Norman, et al [21] was re-examined by O'Keefe, et al [26] who reported that a small annual variation could be detected if the amplitude and phase were allowed to be fitting parameters, rather than constrained to have the same amplitude and phase of the variation of the earth-sun distance.

It has also been observed that the isotopes that exhibit this variation have $\beta$ decay in their decay chain, whereas isotopes decaying with only $\alpha$ decay have so far not shown the effect [13], [27]. However, experiments looking for an analogous influence of antineutrinos on $\beta+$ decay by observing decay rates when a nearby reactor was either on or off obtained negative results [28], [29].

Additional evidence of a solar influence includes observed correlations between decay rates during a solar storm [30], [31] , and the detection of other frequency components that could be interpreted as originating from solar processes [7], [12], [14], [15], [17], [32].

In contrast, other searches for the influence of solar flares [20], [33], an eclipse [34], or other frequency components [25] have yielded negative results.

A recent discussion of three prominent explanations of the observations (environmental influences, solar neutrinos, and cosmic neutrinos) has been presented by Sturrock, et al [35]. They observe that one possibility potentially contributing to the conflicting results is that the effect is variable, and may be more easily observed during some time periods than others.

The observed annual variation in activity along with the hypothesis that the solar neutrino flux may be responsible, suggests a tantalizing possibility for the realization of a compact, relatively inexpensive neutrino detector. Even if neutrinos are not involved, no mechanism has been proposed that satisfactorily explains all of the observations. Consequently, explaining the origin of these time variations should lead at the very least to a better understanding of the measurement of radioactivity.

In this work we report the analysis of Ra-226 decay data from eight high-purity germanium (HPGe) gamma ray detectors, labeled for convenience by their nominal original efficiency. These data were taken for routine quality assurance purposes at the NC State PULSTAR reactor laboratory, and consist of detector counts from a standard Ra-226 sample. The data cover a span of about 18 years.

## 2. Description of Data Sets and Hardware

Four of the eight detectors are still in operation, but available information was sparse on the detectors that are no longer available. Table 1 summarizes the information that is available on the detectors and electronics used for the various measurements. The detectors and associated electronics are/were located in typical air-conditioned laboratories. All of the detectors except the 65% detector were located in the same room. The 65% detector is located in a nearby room on the same corridor. Detectors were refurbished on several occasions, and the relative efficiency was measured after the detectors were serviced. These measurements along with their dates are given in column 4 of Table 1.

**Table 1. System Hardware used to collect the data. NA indicates the records are not available. The average 5 minute background counts in the last column were obtained from 20 hour measurements with no Ra-226 sample, and are corrected for the Compton platform. The background counts should be compared to the approximate mean counts with the sample in place, as shown in Figure 1.**

| Nominal Efficiency | Model | Present MCA Configuration | Measured Efficiency* (Date) | Shield thickness | 609 keV 5 min background, Feb 2014 |
|---|---|---|---|---|---|
| 21% | NA | NA | NA | NA | NA |
| 23% | GEM-20190 | NA | 21.2% (25 Nov 1986) | NA | NA |
| 24% | NA | NA | NA | NA | NA |
| 25% | Ortec GEM-15190-P | AFT 2025 Research amp, Canberra Multiport 2, 3105 HV Supply† | 23.2% (25 Apr 1996) 24% (31 Jan 2008) 21.3% (28 June, 2010) | 4 in (101.6 mm) | NA |
| 26% | NA | NA | NA | NA | NA |
| 38% | Ortec GEM-35190 | AFT 2025 Research amp, Canberra Multiport 2, 3105 HV Supply† | 39.7% (1 Aug 2001) 38.3% (11 Aug 2003) 35.1% (5 May 2005) | 2 in (50.8 mm) | 8 (~0.03%) |
| 42% | Ortec GEM-40195 | AFT 2025 Research amp, Canberra Multiport 2, 3105 HV Supply† | 38.7% (8 Aug 2003) 40.3% (14 Mar. 2011) | 2 in (50.8 mm) | 10 (~0.04%) |
| 65% | Canberra GC6519 | AFT 2025 Research amp, Canberra Multiport 2, 3105 HV Supply | 65.0% (10 Sept 1991) 65% (20 Dec 2005) | 8 in (203.2 mm) | 9 (~0.02%) |

*Relative efficiency at 1.33 MeV, $^{60}$Co
†The MCA electronics for the 25%, 38%, and 42% detectors were originally composed of Ortec components, but were later converted to Canberra. The dates of the conversions are not known.

The calibration source was a ~2 microcurie sealed source of Ra-226 placed ~6 cm above each detector inside a lead shield. Ra-226 decays to Radon-222 via alpha decay, then to Po-214 via two more alpha decays and two beta decays before a final alpha decay to Pb-210. Gamma rays at 609.3 keV are emitted from the first excited state of Po-214. The data sets include the counts in the Po-214, 609.3 keV peak, corrected for the Compton scattering background. The detector resolution, about 2 keV, was also recorded. The counting time was five minutes (live time), and dead times were less than ~8%. The frequency of the measurements typically ranged from 1 to 3 times per week, although there are occasional periods of a week or more where no measurements were taken. The same source was used for all measurements over the entire time period.

Since the decay of radon also produces gamma rays at this energy, it is desirable to know what fraction of the 609.3 keV counts is caused by environmental background instead of the Ra-226 sample. Environmental background measurements for the data sets are not available, but measurements for 3 of the detectors presently in operation were made in February 2014 for reference. The results of these measurements are given in the last column of Table 1. Additionally, radon levels in the basement labs at Burlington Hall where the data were taken were measured to be low (approximately 7 Bq per cubic meter).

The data were recorded by hand on log sheets, so that transcription into Excel spreadsheets was required for analysis. After transcription, outliers were examined manually to eliminate transcription errors. The Excel data were subsequently imported to MATLAB$^{TM}$ for detailed data analysis. The count data sets (corrected for dead time and the Compton pedestal) are shown in Figure 1 along a common time line. The remaining outliers are likely the result of recording errors and cannot be corrected. Although measurements were not simultaneously taken on multiple detectors, most multi-week time intervals are covered by multiple data sets. The descriptions of the unprocessed data sets are given in Table 2.

Individual plots of the data sets are shown in Figure 2. Ordinarily, one would expect the counts to decrease slightly with time, owing to the gradual deterioration of the detectors, and the slight decay of the source (Ra-226 has a half-life of about 1600 years). However, the plots in Figure 2 show significant non-monotonic behavior. The origin of this slow variation of the baseline is unknown, but the variation is generally slow on the scale of the annual cycles that we are primarily concerned with here.

Over time, the energy resolution and to some extent the efficiency of a HPGe detector degrade owing to radiation-induced crystal dislocations and trapped particles. Annealing the detector can help to restore the performance. In some cases, a small amount of the detector surface must be etched away owing to surface contamination. In these cases, the reduction in the detector volume can result in a small reduction in efficiency, though the energy resolution is restored. The dates corresponding to the detectors being serviced are shown by arrows in Figure 2. In the case of the 25% detector, no obvious features in the data are associated with these dates. Similarly, there is a relatively long gap in the 65% data during which the servicing takes place, but no significant change in behavior after the measurements resume. This is consistent with the lack of change in the measured efficiency in December 2005 for the 65% detector.

In contrast, there is a sharp discontinuity after the August 2001 service date for the 38% detector, but no significant features for the August 2003 and May 2005 dates. The lack of a step in May 2005 is particularly surprising, since the change in measured efficiency dropped by more than 8% from the August 2003 measurement.

Discontinuities in the counts are apparent for both service dates in the case of the 42% detector. However, the efficiency measured in March 2011 is about 4% larger than that measured in August 2003, so one would expect the step to go in the opposite direction! There is also what appears to be a small dip in the counts around March 2005. However, close inspection of the behavior here shows that there is no step, but rather a rapid change in the data. Consequently, we take this to be an actual feature in the data.

Unfortunately, we have no way of resolving the apparent lack of consistency between the step sizes and the changes in efficiency measured many years in the past. Consequently, we take the pragmatic approach of attempting to remove the discontinuities that correspond to service dates by data scaling obtained from the local means on either side of the discontinuity. This will be described in more detail in the context of the 38% and 42% detector data sets, and is similar to an approach used by Schrader [36].

In most cases, the data sets were further cleaned to remove remaining outliers, and to omit certain time spans with sparse measurements. The details of the data sets as analyzed are given in Table 3.

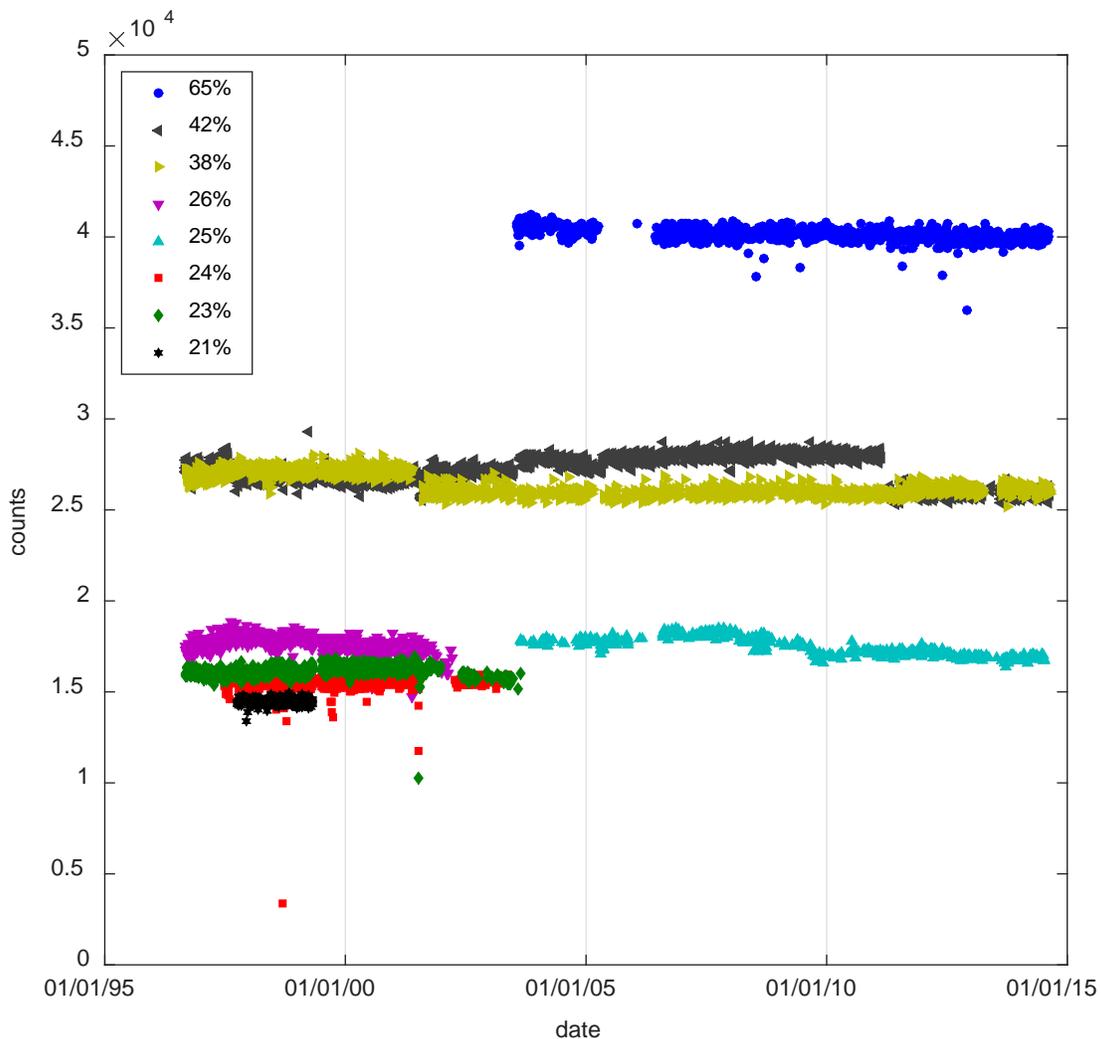

**Figure 1. Raw data from the eight detectors along a common time line. The format of the dates is DD/MM/YY.**

**Table 2. Descriptions of raw data sets from PULSTAR quality assurance measurements.**

| Nominal Efficiency | Raw Data Set Start Date | Raw Data Set Stop Date | Raw Data Set Size | Duration (days) |
|---|---|---|---|---|
| 21% | 6 Oct 1997 | 28 Apr 1999 | 337 | 569 |
| 23% | 3 Sept 1996 | 21 Aug 2003 | 1149 | 2543 |
| 24% | 1 July 1997 | 5 June 2003 | 862 | 2165 |
| 25% | 21 Aug 2003 | 9 July 2014 | 415 | 3975 |
| 26% | 3 Sept 1996 | 19 Mar 2002 | 1068 | 2023 |
| 38% | 3 Sept 1996 | 20 Aug 2014 | 2315 | 6560 |
| 42% | 3 Sept 1996 | 20 Aug 2014 | 2661 | 6560 |
| 65% | 28 July 2003 | 11 Aug 2014 | 853 | 4032 |

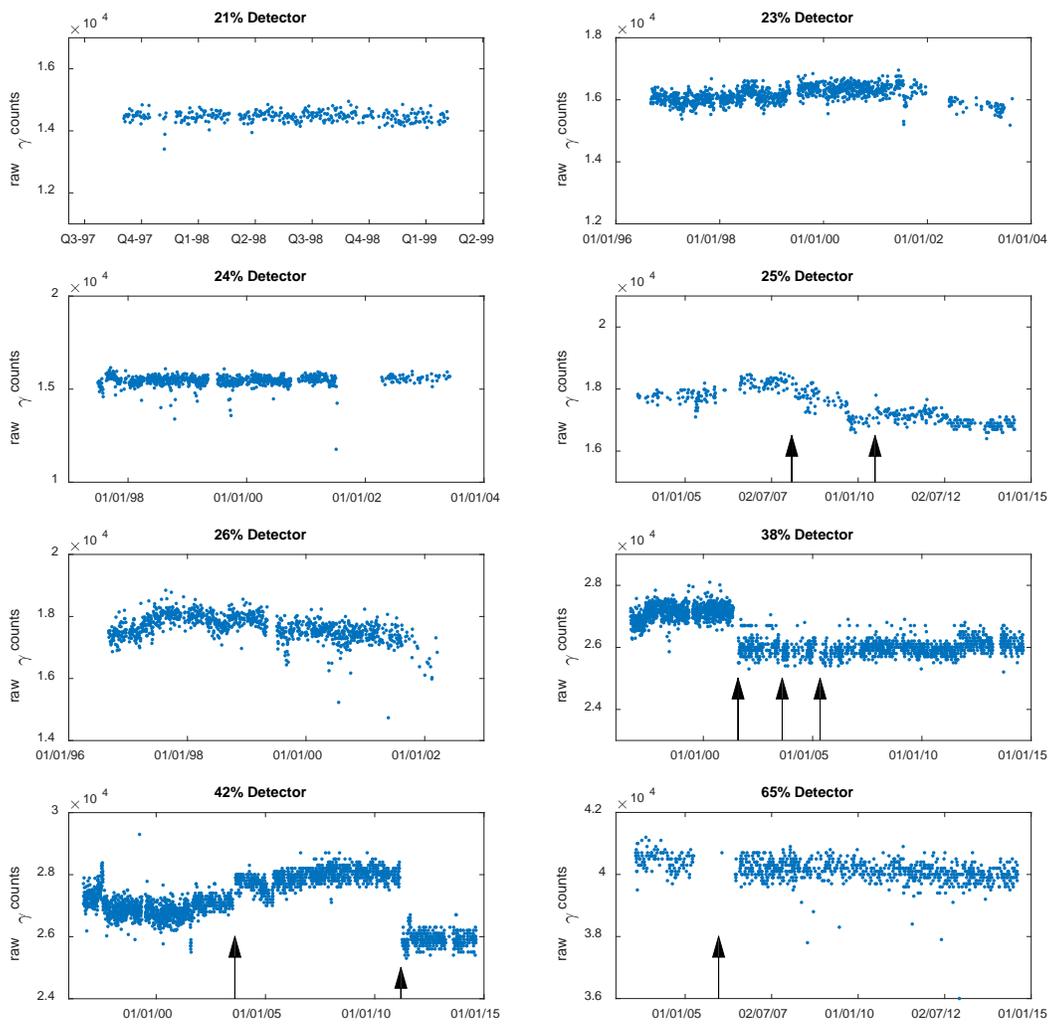

**Figure 2. Individual plots of the complete, unprocessed data sets. The vertical axes are raw gamma counts, and the horizontal axes are dates in DD/MM/YY format. The arrows indicate dates on which the detectors were known to have been serviced.**

**Table 3. Data set dates and sizes as analyzed**

| Detector | Begin Date | End Date | Number of points | Duration (days) | Points in moving average | No. Fourier Components Removed ($f_{low}$ = 0.5 cpy)† |
|---|---|---|---|---|---|---|
| 21% | 6 Oct 1997 | 28 Apr 1999 | 333 | 569 | 18 | 0 |
| 23% | 3 Sep 1996 | 20 Dec 2001 | 1086 | 1934 | 17 | 2 |
| 24% | 1 July 1997 | 6 July 2001 | 783 | 1466 | 17 | 2 |
| 25% | 21 Aug 2003 | 09 Jul 2014 | 414 | 3975 | 4 | 5 |
| 26%-A | 3 Sep 1996 | 19 Mar 2002 | 1054 | 2023 | 16 | 2 |
| 26%-B | 19 Aug 1997 | 19 Mar 2002 | 831 | 1673 | 15 | 2 |
| 38% | 29 Aug 1997 | 17 Jun 2011 | 1756 | 5040 | 11 | 6 |
| 42%-A | 3 Sep 1996 | 20 Aug 2014 | 2655 | 6560 | 13 | 8 |
| 42%-B | 5 Aug 1997 | 25 Jul 2001 | 850 | 1450 | 18 | 1 |
| 42%-C | 10 May 2005 | 12 May 2011 | 751 | 2193 | 11 | 3 |
| 65%-A | 14 Jun 2006 | 11 Aug 2014 | 731 | 2980 | 8 | 4 |
| 65%-B | 23 Feb 2009 | 05 Dec 2013 | 404 | 1746 | 14 | 2 |

†cpy = cycles per year

## 3. Spectrum Analysis of Data Sets

The data set that has the shortest time duration is that from the 21% detector. We begin by describing the detailed analysis of this data set.

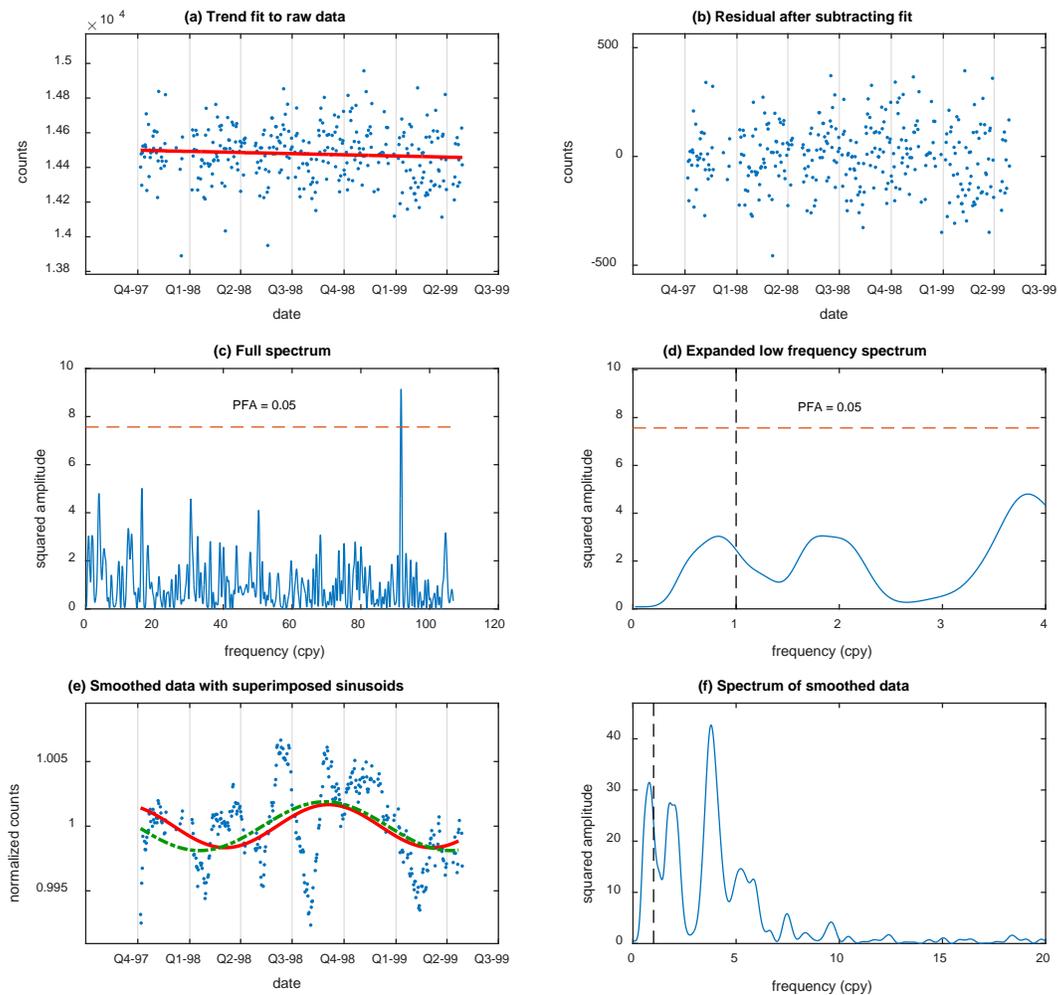

**Figure 3. Analysis of data from the 21% detector. (a) Linear fit to data after removing outliers and correcting for decay. (b) Counts residual after subtracting the linear fit and removing remaining outliers. (c) Complete periodogram. (d) Periodogram expanded for low frequencies. (e ) Data smoothed with a moving average, with sinusoidal fits superimposed. The solid (red) curve is the component with frequency 1 cpy, and the dashed (green) curve is for the frequency of the peak closest to 1 cpy. (f) Low frequency periodogram of the smoothed data.**

### 3.1 Analysis of 21% Detector Data

The analysis of the 21% detector data is shown in Figure 3. First, the outlier near 13,400 counts was removed (see Figure 2(a)), a correction was made for the decay of Ra-226 using a lifetime of 1600 years, and a first-order polynomial was fit to the corrected data. The cleaned and corrected data are compared with the polynomial fit in Figure 3(a). Note that there is still a downward slope after the decay correction. This may be partially due to degradation of the detector, but likely has other contributions of unknown origin.

After subtracting the linear fit, the data set was then clipped a second time to remove points beyond ±3 standard deviations, eliminating the lowest two remaining data outliers. The outlying data points were removed owing to the suspicion that there was an error in the hand-recorded data. (In reality, however, the presence of such a small number of outlying points would have a negligible effect on the analysis.) Finally, the mean was reset to zero after the clip by subtracting the new mean. The result is shown in Figure 3(b). Spectrum analysis was then performed on this data set. Since the time unit of the raw data is days, conversion to years was performed using the factor 365.24219 days/year appropriate for the duration of the tropical year.

The frequency analysis was performed using Lomb-Scargle [37], [38] analysis. Least-mean-square (LMS) fitting of sinusoids was used to obtain the amplitude and phase of specific components, and the Scargle [38] formulation was used to construct periodograms. The complete periodogram is shown in Figure 3(c). The highest frequency computed was determined from the reciprocal of the average time interval between measurements as $N/(2T)$, where $N$ is the number of measurements and $T$ is the duration of the data set.

The probability of the null hypothesis, i.e., the hypothesis that no signal component is present at a given frequency, can be estimated using the method of Scargle [38]. Specifically,

$$PFA = 1 - \left(1 - e^{-P}\right)^M , \qquad (1)$$

where $PFA$ is the probability of false alarm, $P$ is the squared amplitude of the spectrum, and $M$ is the number of independent frequencies. For the present calculations, the value of $M$ was estimated by the number of peaks in the spectrum up to the average Nyquist frequency for each data set [39]. The threshold for a Probability of False Alarm of 0.05 is indicated by the horizontal dashed lines in Figure 3(c,d). Peaks extending above this threshold have a $PFA$ of less than 0.05 and are considered to be statistically significant. The only peak exceeding this threshold occurs at a frequency of 91.7 cycles per year (cpy). This corresponds to a period of about 4 days, but the origin is unknown.

An expansion of the low-frequency portion of the periodogram is shown in Figure 3(d). The vertical dashed line indicates the frequency 1 cpy. The nearest peak is slightly lower at a frequency of 0.82 cpy. Note that the resolution estimated by $1/T = 0.64$ cpy is a measure of the width of the peaks, but does not reflect an uncertainty in the frequency of the peak. The uncertainty in the frequency of a single sinusoid embedded in noise has been discussed by [40], [41], and is estimated by

$$\Delta f = \frac{3\sigma_N}{4TA\sqrt{N}} , \qquad (2)$$

where $\sigma_N$ is the standard deviation of the noise after subtracting the signal and $A$ is the amplitude of the signal. In calculating this uncertainty, the method of Ferraz-Mello [42] was used for the subtraction to minimize spectral leakage arising from non-uniform time samples. The frequency uncertainty of the 0.82 cpy peak was found to be 0.14 cpy. This uncertainty is large enough to make the presence of a 1 cpy component at least plausible. However, since there may be multiple frequencies present in the data set, Eqn. (2) may not be rigorously applicable. Consequently, this and the frequency uncertainty measurements for the other data sets should be interpreted with caution.

Next, the data from Figure 3(b) were low-pass filtered using a centered moving average, and normalized to the mean of the original data (the only processing performed prior to taking this mean was correcting for sample decay). The duration of the moving average is about 30 days. (The actual length of the average was determined by calculating the average number of records per day, and then multiplying by 30. The length of the moving average had a uniform number of records in the average, but because of the uneven temporal spacing, the corresponding number of days varies somewhat over the average.) The result is shown in Figure 3(e). Also shown by the solid line (red) is the LMS best fit in amplitude and phase for a frequency of 1 cpy, and shown by the dashed line (green) is the LMS best fit in amplitude and phase for a sinusoid of frequency 0.82 cpy, the actual peak closest to 1 cpy. Both sinusoidal curves were constructed using the fits to the unsmoothed data. Note that the best fit sinusoids have their minima near the beginning of the year.

As a check, it is interesting to examine the periodogram constructed from the smoothed data. Referring to Figure 3(f), note that frequencies above about 12 cpy are significantly attenuated, as expected from smoothing with an average of about 30 days. Comparison between Figure 3(d) and Figure 3(f) shows that the differences in the periodograms for frequencies below 5 cpy are minor.

For the 21% detector unsmoothed data, the probability of the null hypothesis differs negligibly from 1 for both a frequency of 1 cpy, and the peak nearest 1 cpy. Note that this estimate assumes a single sinusoid in white Gaussian noise, and the validity of these assumptions has not been established. The parameters from the Lomb-Scargle analysis and the LMS fits for a frequency of 1 cpy are summarized in Table 4, and parameters of the peaks closest to 1 cpy are summarized in Table 5. As we will discuss, these tables show that statistically significant frequency components at 1 cpy were observed in four of the eight data sets.

**Table 4. Summary of LMS fits to the detector data sets for a frequency of 1 cpy**

| Detector | Normalized Amplitude | Actual Amplitude (counts) | Phase (days from Jan 1)† | Prob. of Null Hypothesis |
|---|---|---|---|---|
| 21% | 0.00167 | 24.1 | -117.6 | 1 |
| 23% | 0.00333 | 54 | -141.7 | 2.5E-7 |
| 24% | 0.00142 | 21.9 | -17.5 | 1 |
| 25% | 0.00094 | 16.5 | -123.2 | 1 |
| 26%-A | 0.00347 | 61.4 | 42.1 | 0.00015 |
| 26%-B | 0.00584 | 103 | 35.6 | 4.2E-14 |
| 38% | 0.000714 | 19.4 | 8.5 | 1 |
| 42%-A | 0.00172 | 46.9 | -142.2 | 3.8E-8 |
| 42%-B | 0.00269 | 72.2 | -144.5 | 1.6E-7 |
| 42%-C | 0.000623 | 17 | -26.0 | 1 |
| 65%-A | 0.00135 | 54.4 | 23.3 | 0.08 |
| 65%-B | 0.00183 | 73.6 | 32.8 | 0.045 |

†Positive values correspond to peaks occurring after 1 January.

**Table 5. Properties of peaks nearest 1 cpy**

| Detector | Frequency (cpy) | Frequency uncertainty (cpy) | Normalized Amplitude | Prob. of Null Hypothesis |
|---|---|---|---|---|
| 21% | 0.822 | 0.14 | 0.00189 | 0.99 |
| 23% | 1.104 | 0.01 | 0.00478 | 3.1E-17 |
| 24% | 0.895 | 0.028 | 0.00299 | 0.0049 |
| 25% | 1.062 | 0.014 | 0.00267 | 0.27 |
| 26%-A | 1.117 | 0.018 | 0.00352 | 0.00013 |
| 26%-B | 0.914 | 0.011 | 0.00668 | 5.3E-18 |
| 38% | 1.003 | 0.014 | 0.000718 | 1 |
| 42%-A | 0.976 | 0.003 | 0.00239 | 1.2E-17 |
| 42%-B | 1.078 | 0.019 | 0.00289 | 2E-9 |
| 42%-C | 0.900 | 0.03 | 0.00107 | 0.96 |
| 65%-A | 0.961 | 0.014 | 0.00157 | 0.0061 |
| 65%-B | 0.961 | 0.026 | 0.00193 | 0.013 |

## 3.2 Analysis of 23% Detector Data

Referring to Figure 2, we note that there is a long-term variation in the 23% detector data that is more complicated than the linear trend of the 21% detector. A higher-order polynomial fit could be used to subtract this trend, but to minimize periodogram distortion from spectral leakage, a fit using orthonormal basis functions based on low-frequency sinusoids was used instead. The frequencies used were multiples of the fundamental Fourier frequency $1/T$ up to a specified frequency $f_{low}$. The method of Ferraz-Mello [42] was then used to subtract out these frequency components with amplitudes and phases determined by LMS fits. This method uses the Gram-Schmidt procedure to obtain orthonormal basis functions based on the set of frequencies to be subtracted. (While not necessary, subtracting out this long-term variation makes it easier to examine the periodogram for annual variations. If this is not done, the low-frequency spectral features from this long-term variation dominate the periodogram. It is also likely that these large amplitude, low-frequency components introduce their own spectral leakage that may be partially removed by this subtraction process.) The analysis of the data from the 23% detector is shown in Figure 4. In this case the sparse data after 2002 were omitted from subsequent analysis. The spectrum does not change significantly if these are included, but the accuracy of the running average used to smooth the data is affected by such a large gap. The outlier at 10,280 counts about midway through 2001 (see Figure 1) was also omitted from analysis. The data corrected for this trend and after removing two additional low-count outliers near mid-2001 are shown in Figure 4(b). The full spectrum in Figure 4(c) shows increased power near the frequency of 52 cpy. This feature is present in several of the data sets, and is likely an artifact associated with a weekly pattern imposed on the measurement intervals by the lack of measurements on weekends. The largest peak has a frequency of 1.104 cpy. In this case, the small frequency uncertainty of 0.01 cpy suggests that the deviation from 1 cpy of this peak is significant. Figure 4(e) shows the data smoothed with a moving average of about 30 days along with the best fit sinusoids at frequencies of 1 cpy (solid/red) and 1.104 cpy (dashed/green). Although they become significantly out of phase with each other over such a long time period, we see that they are in phase during 1998-2000 where the annual component is most pronounced in the smoothed data curve. Again we note that the sinusoids are close to minimum at the beginning of the year during this time period. As before, the parameters for the Lomb-Scargle analysis and the LMS fits are summarized in Table 4 and Table 5.

The fact that annual variations are easier to see in some time periods than others suggests that the frequency components are time varying. Time variations in similar periodograms have been previously reported [9-12], [17], [35]. To better visualize the time variations, a short-time periodogram is shown in Figure 4(f). The periodogram was constructed using a sliding, centered window of width 2 years. Because of this shorter time window, the frequency resolution is much lower than in Figure 4, but some frequency drift as well as a concentration in the late 1998 and early 1999 time frame are apparent.

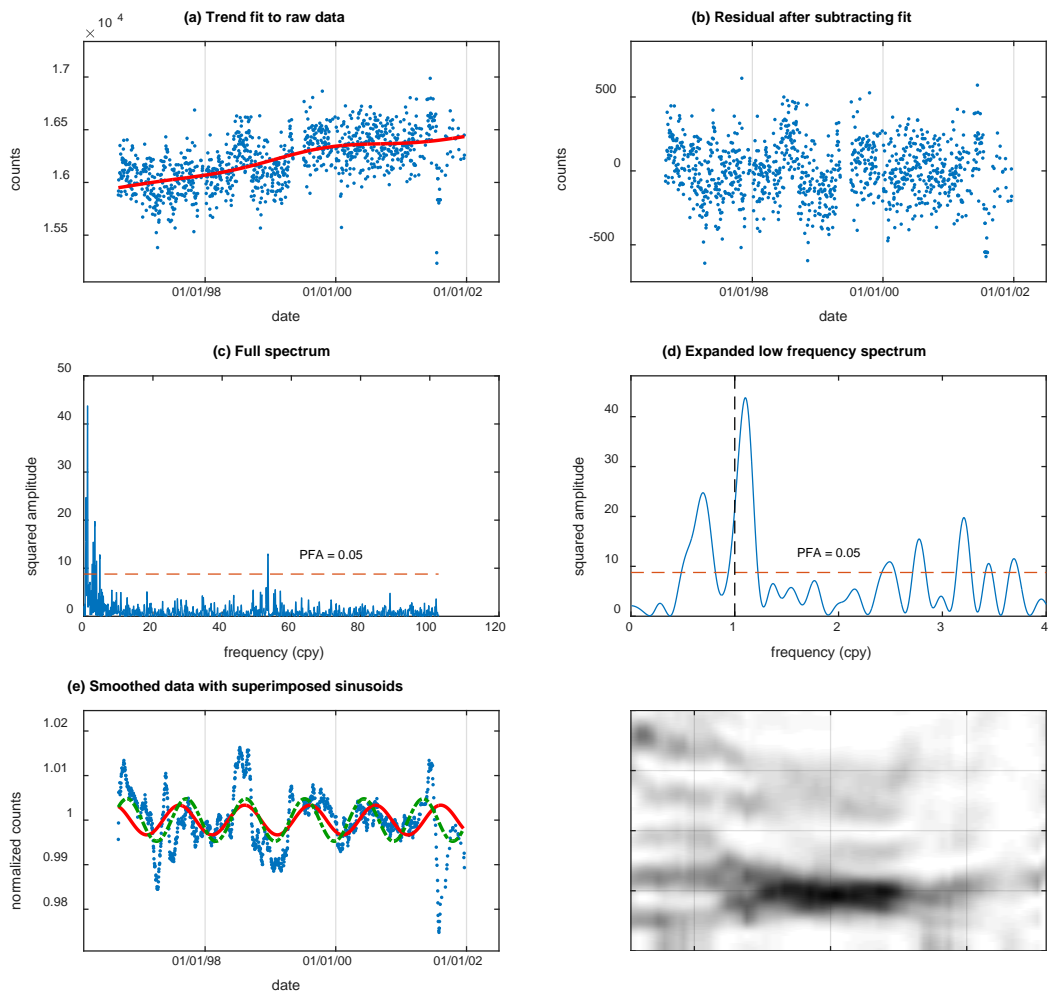

**Figure 4.** Analysis of data from the 23% detector. (a) Trend fit to data after removing outliers and correcting for decay. The two lowest Fourier component frequencies were used in the fit. (b) Counts residual after subtracting the trend fit and removing remaining outliers. (c) Complete periodogram. (d) Periodogram expanded for low frequencies. (e) Data smoothed with a moving average, with sinusoidal fits superimposed. The solid (red) curve is the component with frequency 1 cpy, and the dashed (green) curve is for the frequency of the peak closest to 1 cpy. (f) Short-time periodogram. The annual variations are most apparent during the latter half of 1998 and the first half of 1999.

## 3.3 Analysis of 24% Detector Data

The analysis of the data from the 24% detector is shown in Figure 5. As with the 23% data, the measurements after the gap near the beginning of 2002 were omitted from the analysis. In this case, the peak closest to an annual variation has a frequency of 0.895 cpy. However, the frequency of 1 cpy is located near the null between two adjacent peaks. Consistent with these observations, the plot of the smoothed data in Figure 5(e) does not show any clear annual variation.

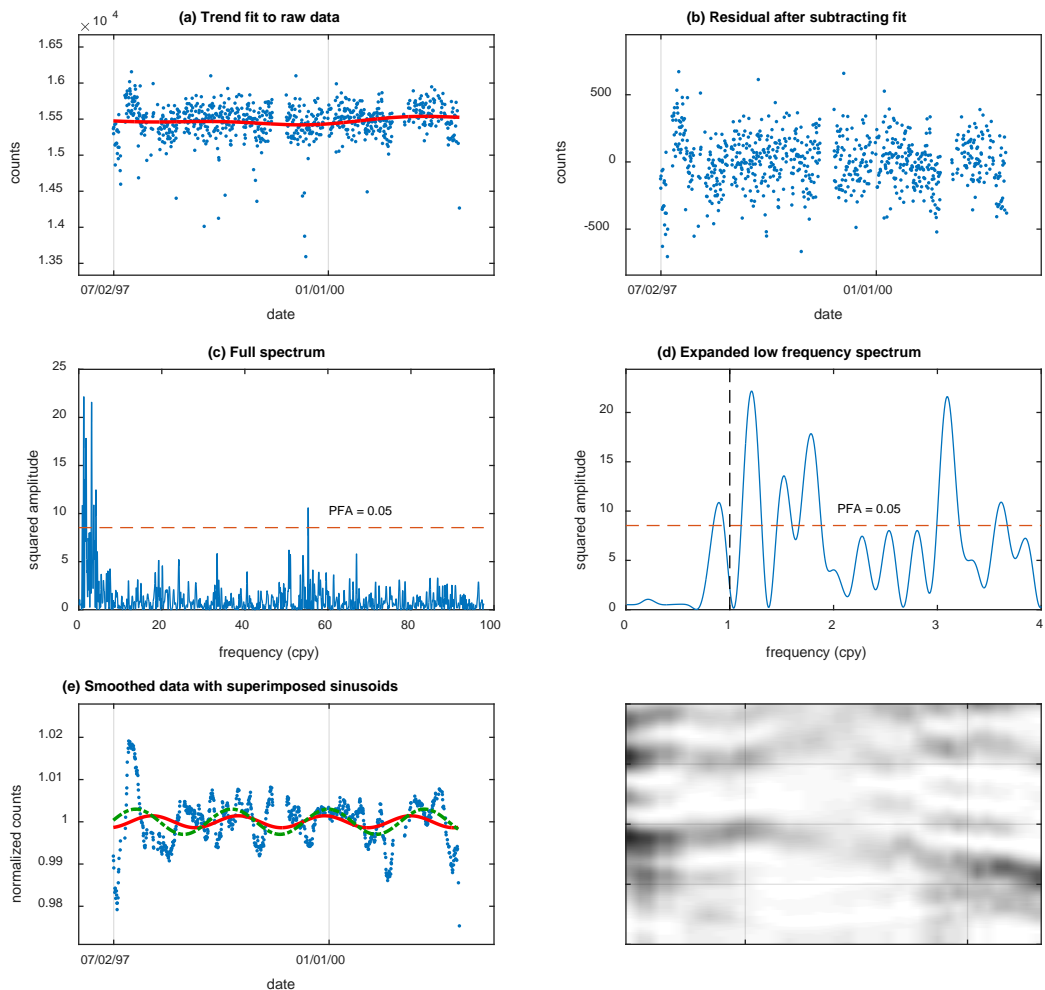

**Figure 5. Analysis of data from the 24% detector. (a) Trend fit to data after removing outliers and correcting for decay. The two lowest Fourier component frequencies were used in the fit. (b) Counts residual after subtracting the trend fit and removing remaining outliers. (c) Complete periodogram. (d) Periodogram expanded for low frequencies. (e) Data smoothed with a moving average, with sinusoidal fits superimposed. The solid (red) curve is the component with frequency 1 cpy, and the dashed (green) curve is for the frequency of the peak closest to 1 cpy. (f) Short-time periodogram.**

## 3.4 Analysis of 25% Detector Data

The data from the 25% detector has a more complex long-term baseline variation than the other data sets, as shown in Figure 6. Five Fourier components were used to subtract out this long-term variation. This data set differs from the previous ones in that beginning in July, 2012, the number of significant digits for the counts that were recorded was reduced from 5 to 3. This quantization can be seen in the data scatter plots in Figure 6(a,b).

The nearest peak shown in Figure 6(d) has a frequency of 1.062 cpy. However, from Table 4, Table 5, and Figure 6, there is no statistically significant component with a frequency of either 1 cpy or 1.062 cpy in this data set.

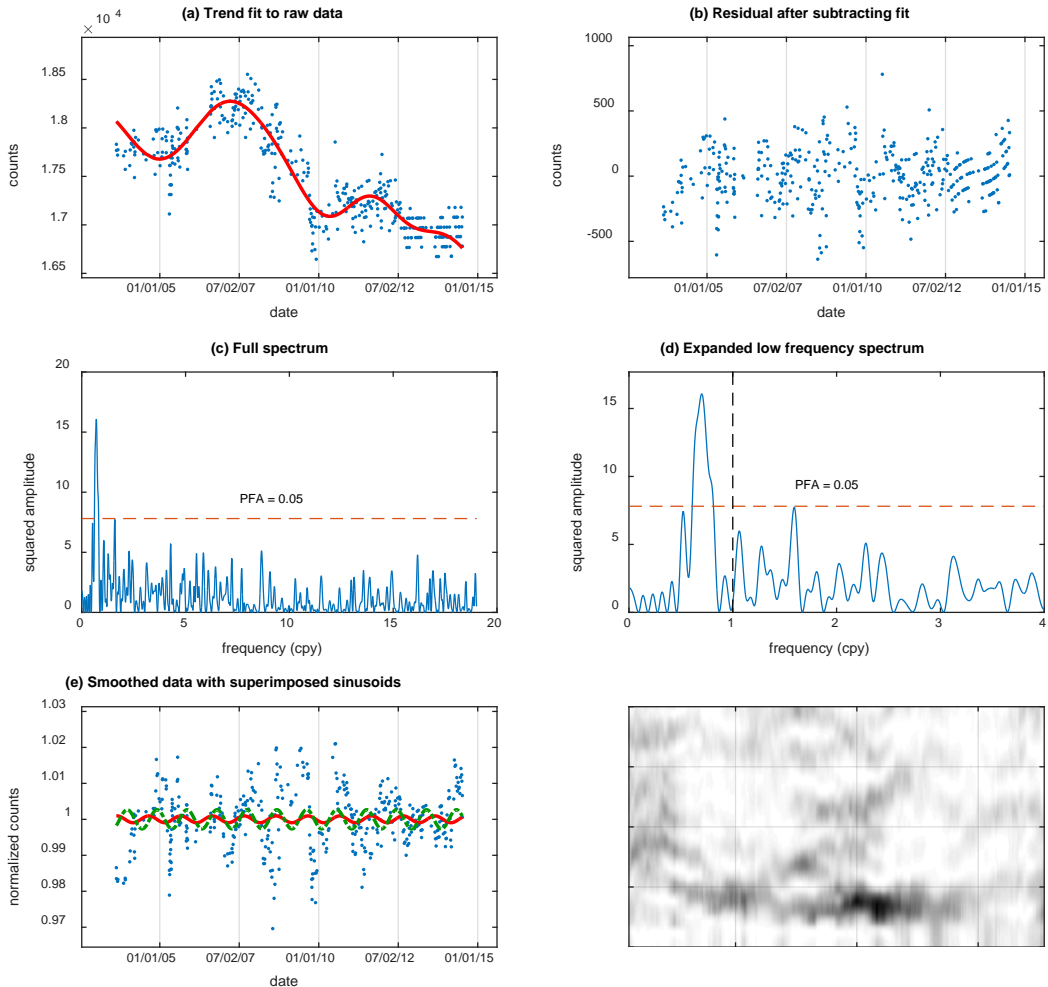

**Figure 6.** Analysis of data from the 25% detector. (a) Trend fit to data after removing outliers and correcting for decay. The 5 lowest Fourier component frequencies were used in the fit. (b) Counts residual after subtracting the trend fit and removing remaining outliers. (c) Complete periodogram. (d) Periodogram expanded for low frequencies. (e) Data smoothed with a moving average, with sinusoidal fits superimposed. The solid (red) curve is the component with frequency 1 cpy, and the dashed (green) curve is for the frequency of the peak closest to 1 cpy. (f) Short-time periodogram.

### 3.5 Analysis of 26% Detector Data

The first 2 Fourier frequencies were used to fit the baseline variation for the 26% detector, as shown in Figure 7(a). The resulting full periodogram shown in Figure 7(c) again shows the small feature near 52 cpy. The largest feature in the unsmoothed data set is a peak with a frequency just below 1 cpy, as shown in Figure 7(d). Examination of the smoothed curve in Figure 7(e) shows that the time interval between minima appears to decrease slightly over the duration of the data set. The solid (red) and dashed (green) sinusoidal curves are closest to being in phase during the latter half of 1998 and the first half of 1999, suggesting that this is the time interval with the strongest component with a period of 1 year. Prior to this the period appears a bit longer, and after this, the period appears a bit shorter. This shift in frequency with time is again shown more clearly in the short-time periodogram shown in Figure 7(e ).

Prompted by this observation, the data set was re-analyzed as shown in Figure 8 where the measurements prior to 19 August, 1997 were omitted. A comparison of Figure 7(d) and Figure 8(d) shows that the peak is indeed shifted closer to the frequency of 1 cpy, and the amplitude of the 1 cpy component is significantly increased.

Note that both the 23% and the 26% data sets show the strongest annual variations during late 1998 and early 1999. However, the 23% data has a minimum near the first of the year, while the 26% data set has a maximum near the beginning of the year, so that the signals are very nearly out of phase between these two data sets.

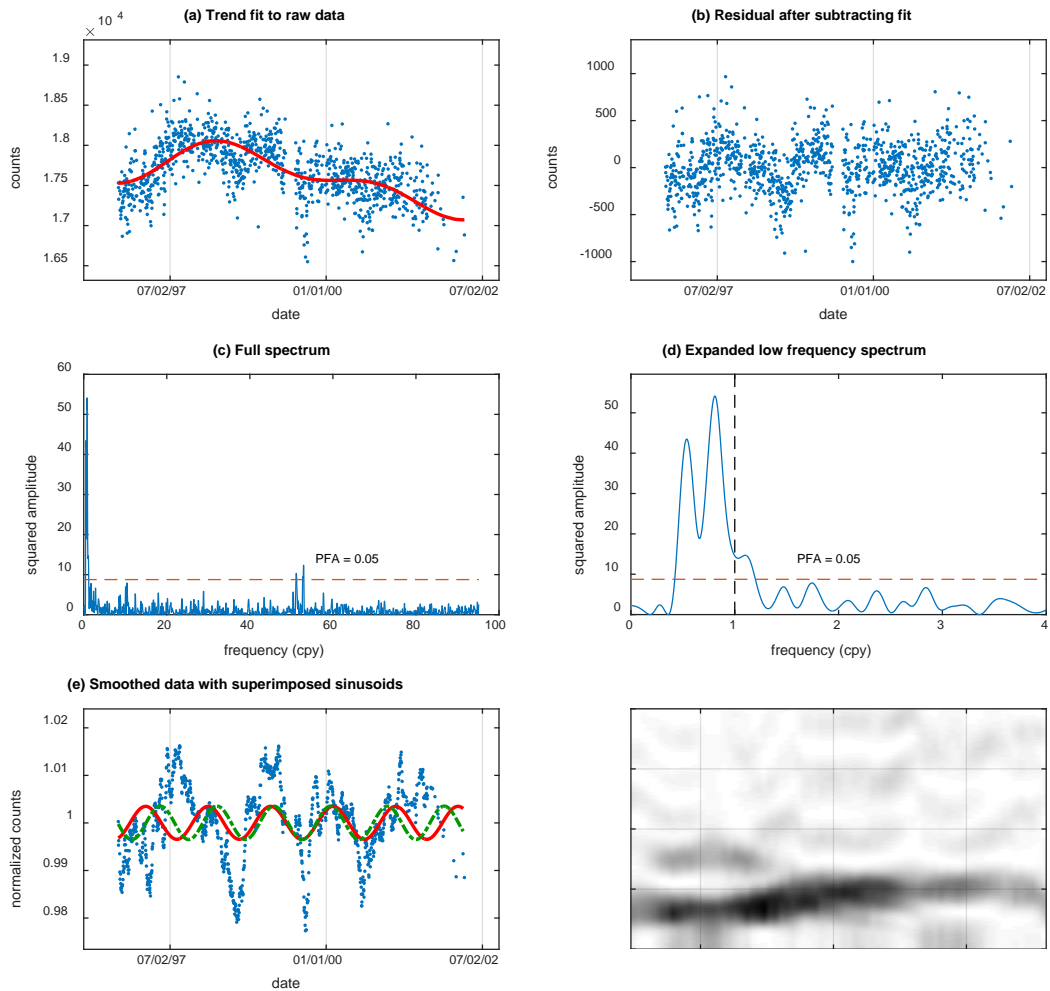

**Figure 7. Analysis of data from the 26% detector. Parameters correspond to row labeled "26%-A" in Table 3. (a) Trend fit to data after removing outliers and correcting for decay. The 2 lowest Fourier component frequencies were used in the fit. (b) Counts residual after subtracting the trend fit and removing remaining outliers. (c) Complete periodogram. (d) Periodogram expanded for low frequencies. (e) Data smoothed with a moving average, with sinusoidal fits superimposed. The solid (red) curve is the component with frequency 1 cpy, and the dashed (green) curve is for the frequency of the peak closest to 1 cpy. (f) Short-time periodogram.**

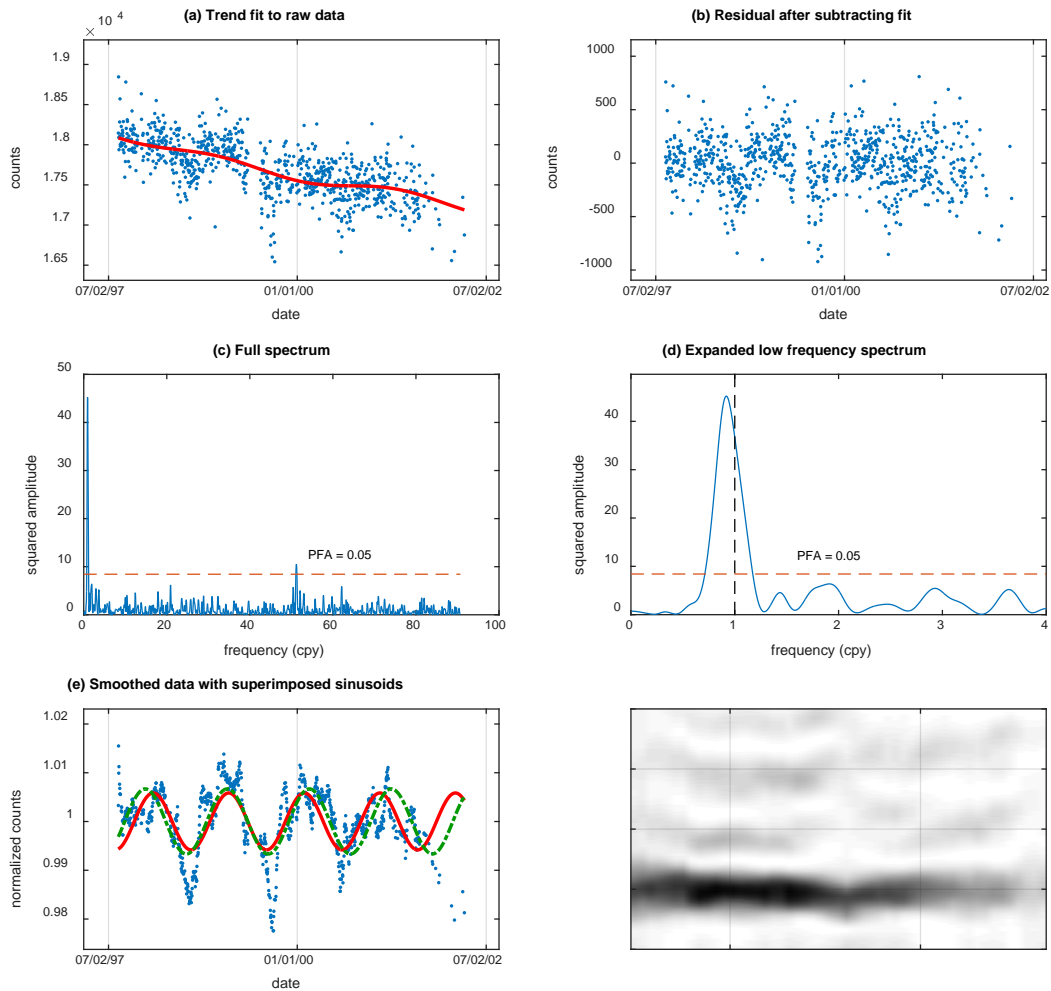

**Figure 8. Analysis of 26% detector data. In contrast with the analysis shown in Figure 7, this analysis begins with 19 August, 1997. The annual variation is particularly noticeable during this time interval. This analysis corresponds to the data set 26%-B.**

### 3.6 Analysis of the 38% Detector Data

Referring to Figure 2, we see that the data from the 38% Detector is the first to have a significant discontinuity corresponding to one of the dates that the detector was refurbished. This discontinuity was removed by averaging the measurements within 1000 days of either side of the discontinuity, and multiplying the values after the discontinuity by the ratio of the means before and after. Again referring to Figure 2 we see that there is a region of increasing activity at the beginning of the set, and what appears to be an unexplained step after June, 2011. To avoid these features, the analysis was limited to the central portion between 29 August 1997 and 17 June 2011. The result is shown in Figure 9(a), along with a fit with 6 Fourier components. Note also that in 2002 the recorded data dropped from a full 5 digits to 3 digits, resulting in the quantization seen in parts (a) and (b).

From Figure 9(c-e) we conclude that there are no statistically significant sinusoidal frequency components.

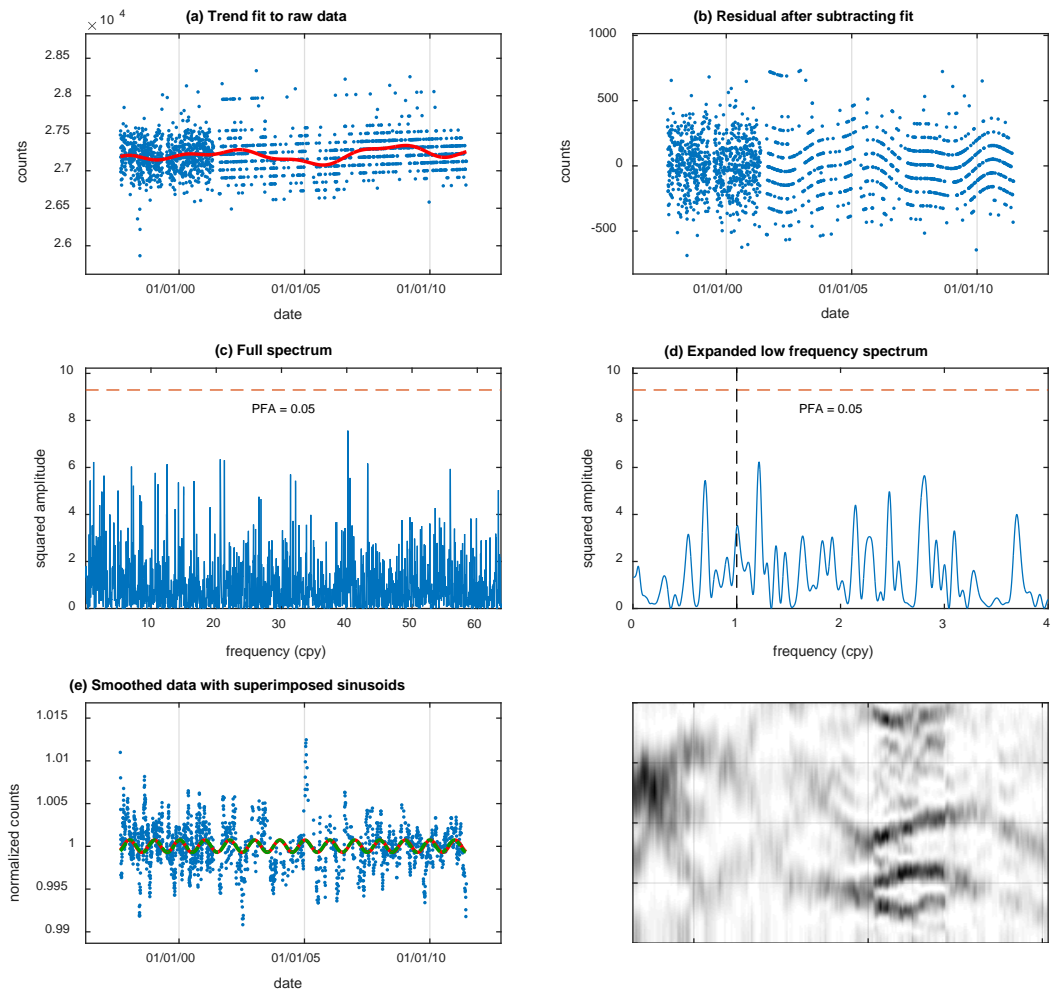

**Figure 9. Analysis of the 38% detector data set.**

### 3.7 Analysis of 42% Detector Data

From Figure 2 we see that there are two discontinuities in the data for the 42% detector, corresponding to dates when the detector was refurbished. As with the 38% detector, these discontinuities were removed prior to analysis by averaging the measurements on either side of the discontinuity, and multiplying the values after the discontinuity by the ratio of the means before and after. For the discontinuity in August, 2003, the means from measurements within 300 days on either side were used, while for the discontinuity in March, 2011, means from measurements within 1000 days on either side were used. The rescaled data along with a fit using the 8 Fourier harmonics with frequencies below 0.5 cpy are shown in Figure 10(a). As with the 38% data, the reduction in recorded digits from 5 to 3 in 2002 results in visible quantization of levels.

Referring to Figure 10(e), we note that the sinusoid at 1 cpy and the sinusoid at the spectral peak nearest to 1 cpy are in-phase prior to about 2004, and the most visible annual variation in the smoothed data is between 1998 and 2001. In contrast, the two sinusoids have become significantly out of phase by 2014, and it is difficult to see any similar variation in the smoothed data after 2006.

We note also that there are 4 "spikes" in the smoothed data whose presence could influence the LMS fits. We do not know if the "spikes" should be considered to be signal or artifact, but in view of the apparent phase slippage over time, it is interesting to separately analyze segments of the data between these spikes. Figure 11 shows the analysis for the time period between the earliest two spikes, 5 August, 1997, through 25 July, 2001 (data set 42%-B in Table 3-Table 5). During this time period a frequency component very near 1 cpy is clearly apparent. For comparison, analysis of the segment between the third and fourth spikes is shown in Figure 12. Although Figure 12(d,e) show a frequency component near 1 cpy, Figure 12(c,d) show that there are many peaks of comparable amplitude, and none reach the threshold for statistical significance. Consequently the likelihood that the peak near 1 cpy is a false alarm is very high, although there is some indication of an annual variation during 2007.

### 3.8 Analysis of 65% Detector Data

As shown in Figure 2, there is a large gap in the data set for the 65% detector during which the detector was refurbished. However, in this case, there does not appear to be a significant change in detector efficiency after refurbishment. As with previous data sets, the data prior to the gap were omitted from the detailed analysis. The results of analyzing the data after the gap are shown in Figure 13. Referring to Figure 13(e), the sinusoid with frequency 1 cpy and that of the closest peak to 1 cpy are most in-phase in the time period around 2010 and 2011. This is also the time period where the "annual" variation is most apparent from the smoothed data. The short-time periodogram is shown in Figure 13(f). Consistent with Figure 13(e), the frequency component near 1 cpy is most pronounced between 2010 and 2012.

The analysis of the data from 2009 through 2013 where the annual variation is most apparent is shown in Figure 14. Over this time period the annual variation can be clearly seen in the smoothed data (Figure 14(b)).

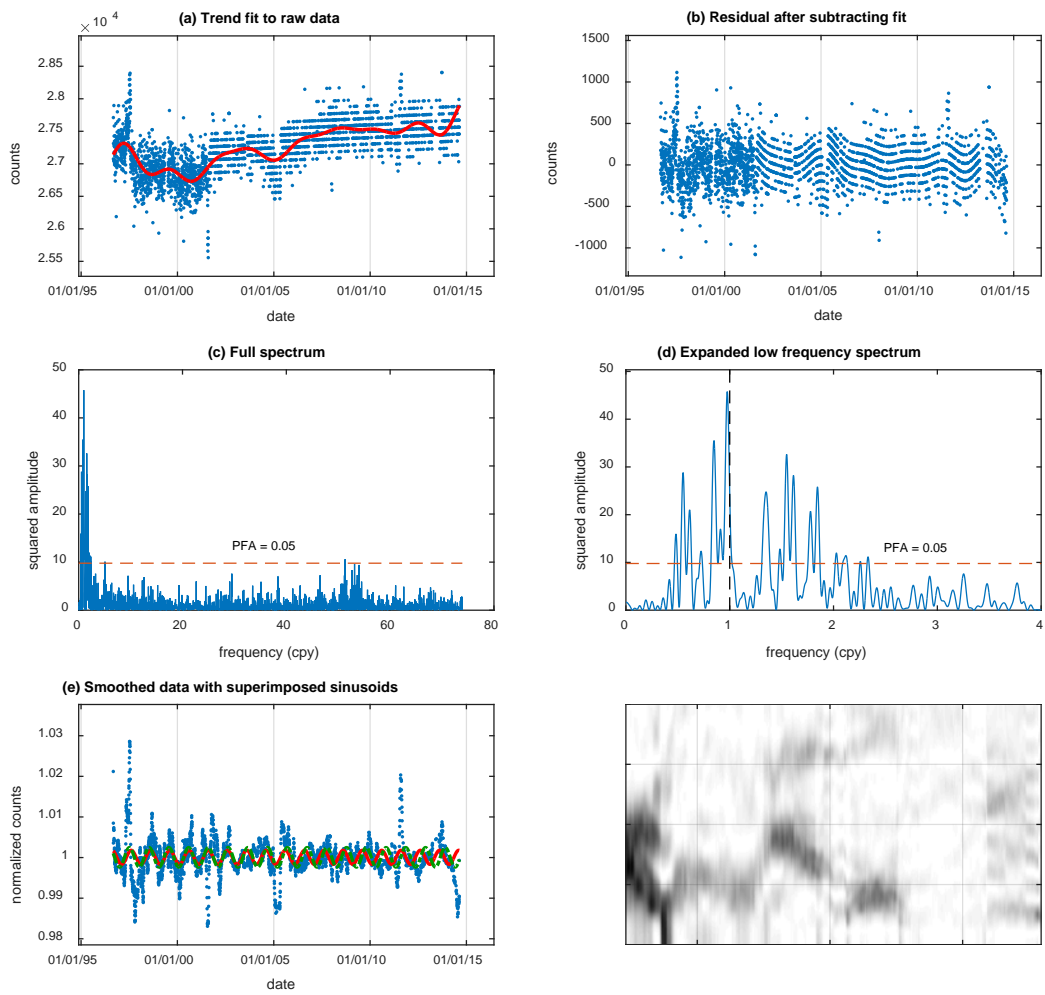

**Figure 10. Analysis of the data from the 42% detector (data set 42%-A in Table 3-Table 5).**

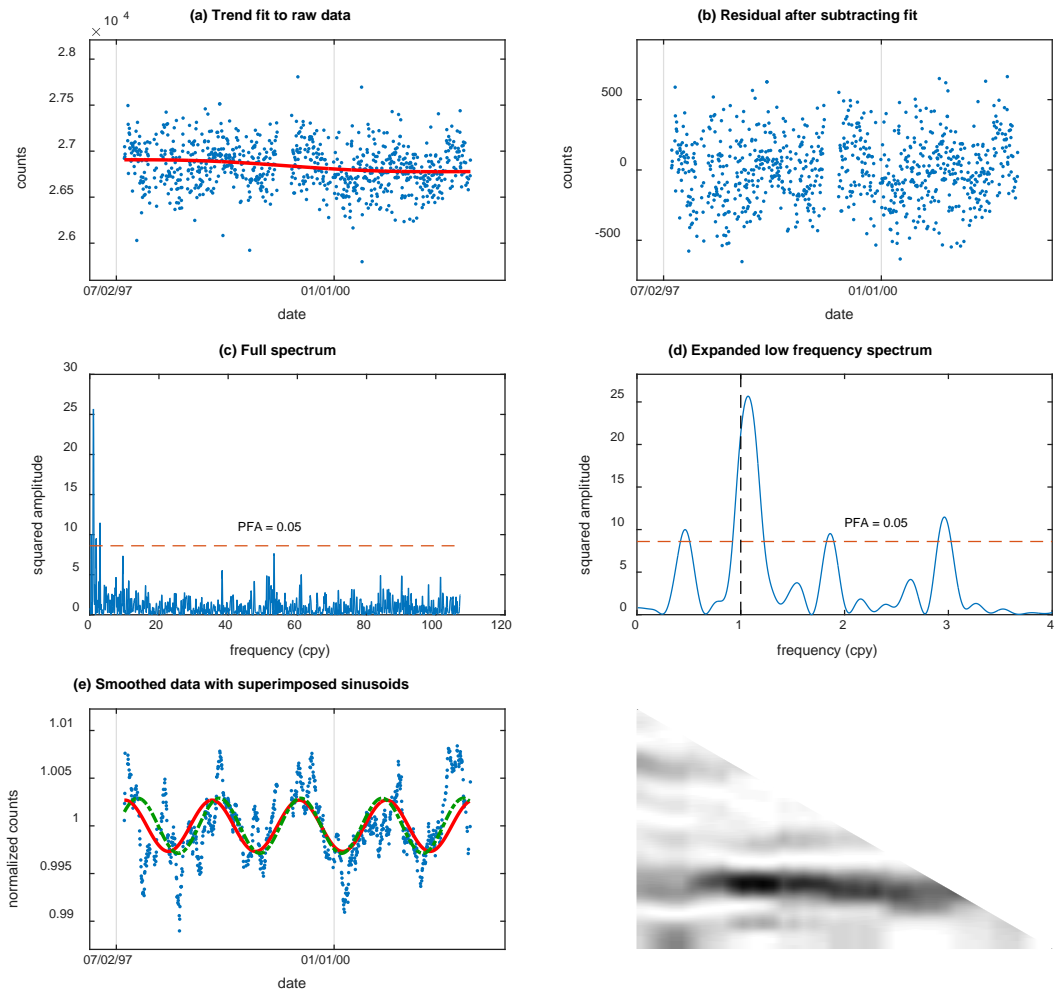

**Figure 11. Analysis of segment of data from the 42% detector between 5 August, 1997, and 25 July, 2001 (data set 42%-B in Table 3-Table 5).**

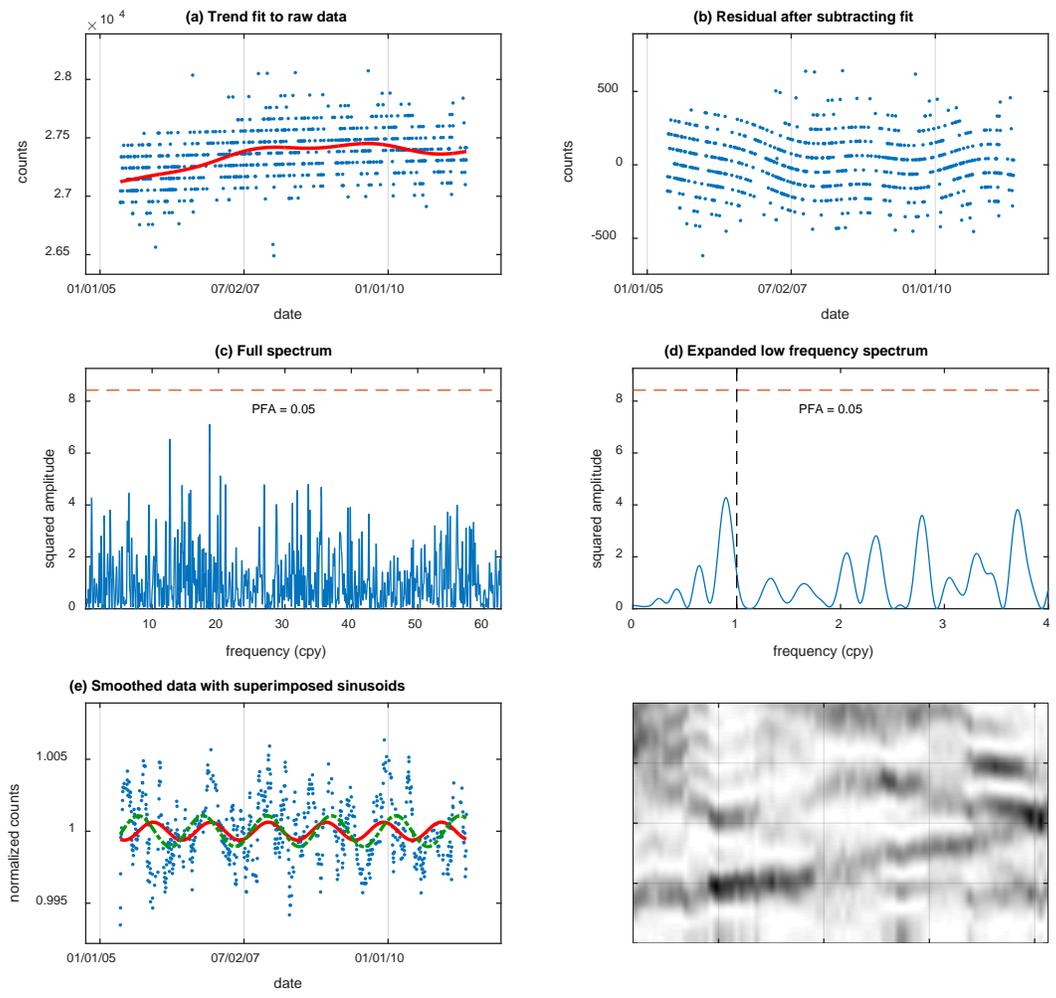

**Figure 12. Analysis of segment of data from the 42% detector between 10 May, 2005, and 12 May, 2011 (data set 42%-C in Table 3-Table 5).**

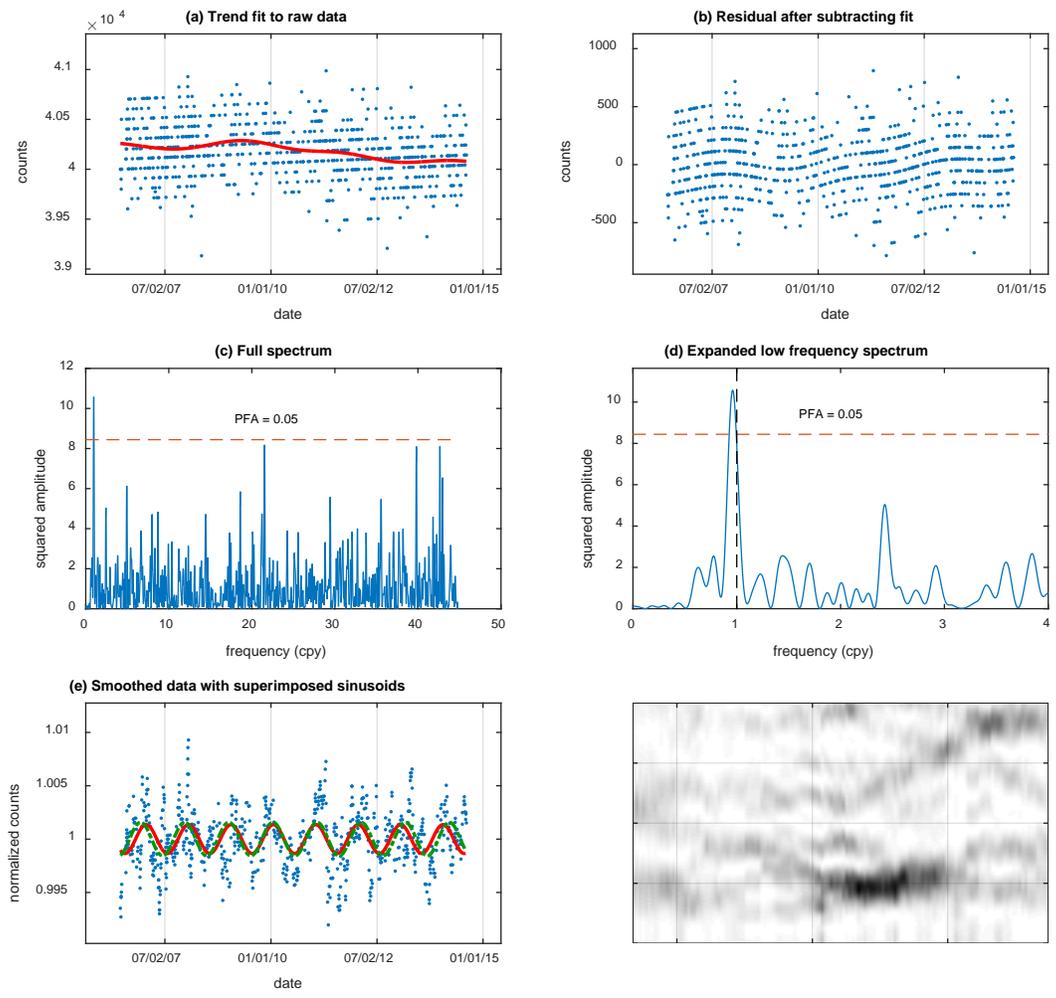

**Figure 13. Analysis of the data from the 65% detector. (65%-A)**

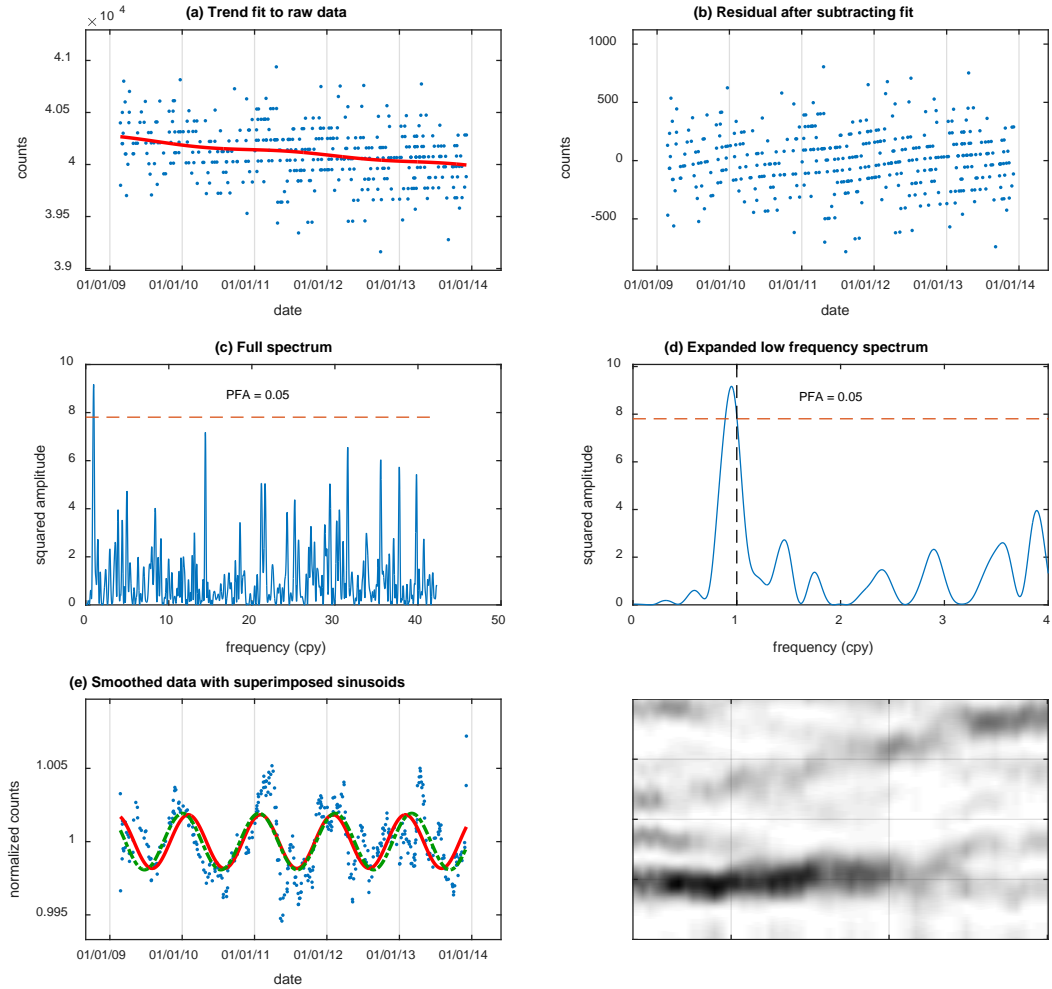

**Figure 14. Analysis of data from the 65% detector during the time period over which the annual variation can be most clearly seen. (65%-B)**

## 4. Analysis of Simulated Data

To gain a better understanding of the accuracy of the analysis procedure when the amplitude of the periodic signal is very small, simulated data sets were constructed and then analyzed using the same procedure used on the PULSTAR data sets. To model the uneven time sampling, measurement dates were generated such that the intervals between measurements were Poisson distributed with the rate $\lambda_{days}$ measurements per day. Similarly, the number of counts in a fixed interval was modeled as a Poisson process with the rate $\lambda_{counts}$ counts per second. The specific parameters used in the simulations are given in Table 6 and were selected to roughly mirror the data from the 65% detector. Six values of normalized signal amplitude were considered, ranging from 0 to 0.2%, and 1000 data sets were simulated for each value of signal amplitude. The results are shown in Figure 15.

Table 6. Parameters used to generate simulated data sets.

| Parameter | Value |
|---|---|
| $\lambda_{day}$ | 0.25 measurements/day |
| Number of measurements | 700 |
| $\lambda_{counts}$ | 133 counts/second |
| Counting interval | 300 seconds |
| Normalized signal amplitudes | 0, 0.025%, 0.05%, 0.1%, 0.15%, 0.2% |
| Delay from 1 January | 100 days |
| Number of cases per signal amplitude value | 1000 |

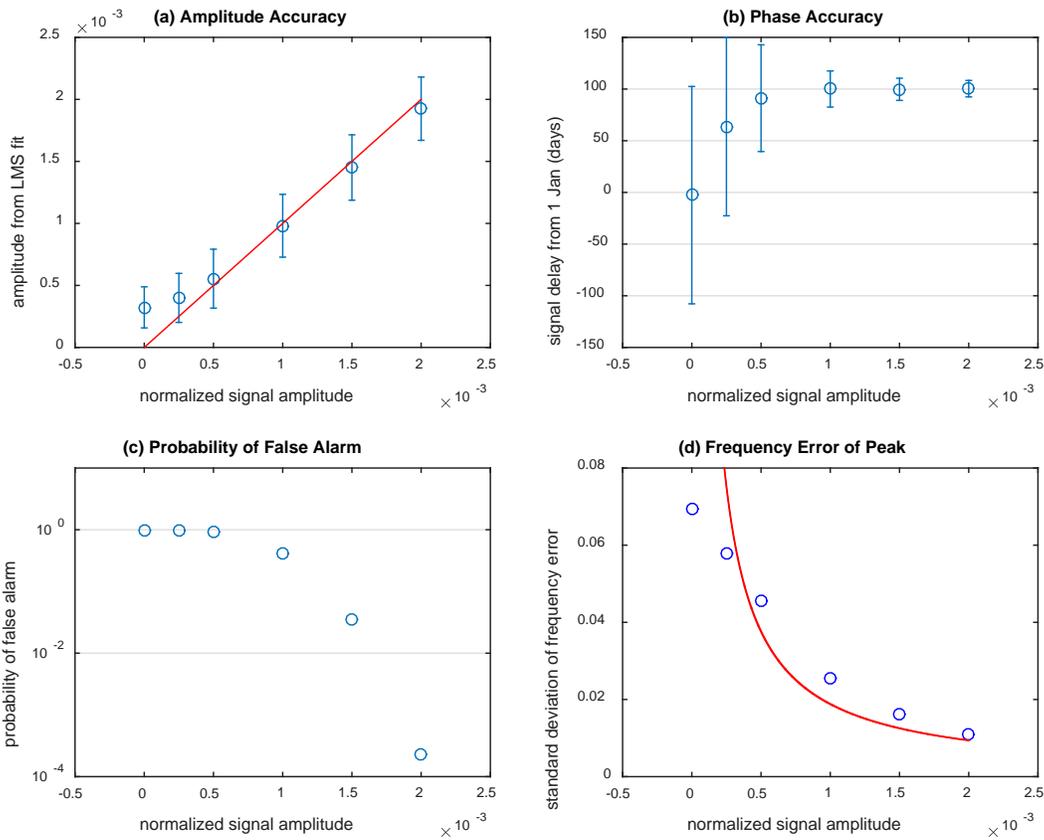

Figure 15. Analysis of simulated data sets. (a) Accuracy of the signal amplitude from the LMS fit. The solid line has unity slope and is included for reference. (b) Accuracy of the signal delay in days from January 1 from the LMS fit. The correct delay was 100 days for each amplitude. (c) Probability of false alarm for each amplitude value. (d) Comparison of the standard deviation of the frequency error from the simulation with Eqn. (2). In each subfigure, each data point is the average of 1000 simulations, and the error bars (when given) represent ± 1 standard deviation.

With no a priori knowledge about the presence or absence of a signal, Figure 15(c) indicates that a normalized amplitude of at least about 0.15% would be required to have confidence that a signal has been detected. However, it is interesting to observe that if one has a priori knowledge that the signal is present, Figure 15(a,b) show that the analysis gives reasonable estimates of the amplitude and phase down to amplitudes of about 0.05%.

Referring to Table 4, we see that the amplitudes of the signals in the data sets with statistically significant components at 1 cpy are larger than 0.15% as expected (i.e., 23%, 26%-A,B, 42%A,B, 65%B). At the same time, the amplitudes in all of the data sets are greater than 0.05%, suggesting that useful phase data may still be present even if the PFA is high. We will return to this point in the next section.

Figure 15(d) compares the standard deviation of the frequency error from the simulation with Eqn. (2), showing good agreement.

## 5. Discussion

Since environmental radon also contributes to gamma rays at 609 keV, it is of interest to examine whether seasonal variations in this source could explain the observed periodicity in the measured counts. Measurements of background taken over the durations of the data sets are not available. However, as mentioned previously, measurements of the ambient 609 keV line were performed in February 2014 for the 38%, 42%, and 65% detectors. Of these detectors, only the 42% and 65% showed statistically-significant components with frequencies of 1 cpy in the analyzed data sets. Owing to the phase of the signal observed from the 42% detector, February should be close to the minimum of the annual component, and we cannot rule out the possibility that counts would increase 6 months later. In contrast, the delay of the 65% signal puts the peak near the beginning of February, so the annual component should be near its maximum. With an estimated amplitude of about 73 counts and recognizing that the counts must be positive definite, the peak value should be twice this, or about 146 counts. In contrast, the measured contribution of background sources translates to an expected 9 counts during the measurement interval. Consequently, at least for this one detector and measurement date, the background contributions to the 609 keV line are more than an order of magnitude too small to explain the observed annual variation.

Referring to Table 4, using the Probability of False Alarm as a semi-quantitative metric, data from only 4 of the 8 detectors show a significant spectral component at a frequency of 1 cpy. These four are also the only ones for which something approximating an annual variation is visually apparent from the smoothed curves of counts versus time. Of these four, the 23%, 26%, and 42% data show the variations over approximately the same time frame between 1997 and 2001. In contrast, the annual variation appears over the interval 2010-2012 for the 65% detector.

Figure 16 shows the smoothed counts versus time curves for the 23%, 26%, and 42% detectors. The curves represent the residuals after subtracting out the means, and normalizing to the appropriate mean. The 26% and 42% curves are shown with a small offset to better enable the comparison between the curves. Close examination of the curves reveals an interesting and puzzling fact (also indicated by the phase data in Table 4): the 42% and 23% curves are near a minimum at the first of each year, while the 26% curve is near maximum at the first of each year. In other words, even though all of the detectors are of the same type, in close physical proximity, and with similar electronics and processed with identical software, the data from the 26% detector is very nearly out of phase with the other two detectors.

The fact that all three contain a similar frequency component over the same time interval is consistent with an external influence on the count rates, but the fact that one is out of phase with the other two suggests that something about the specific detector and/or processing can influence the phase—at least by changing the sign.

Other phase inconsistencies have been reported in the literature. A change in sign was observed by Schrader [36] between the phase of Kr-85 and the ratio of of Kr-85/Eu-154. Also, the phase reported by [36] for Eu-152 using an ionization chamber is approximately inverted from that of Eu-152 by Siegert using a Ge(Li) detector [3].

Steinitz et al [16], [43] reported phase differences depending on the placement of the detector with respect to the source. However, in our case, the sample of Ra-226 was placed in the same location with respect to the detector: directly above and centered with the axis. So a possible emission anisotropy would not explain the differences in phase observed in our data.

A polar plot showing the amplitudes and phases of the signals in the eight data sets is shown in Figure 17. The four data sets with statistically significant components at a frequency of 1 cpy are shown with solid circles. Consistent with the observation from the simulations, we might hypothesize that a signal is also present in the remaining four data sets so that the amplitude and phases may give meaningful estimates even though the statistical probability of false alarm is high. The amplitudes and phases of the remaining data sets that do not show statistically significant signals are shown by the open circles in Figure 17. The open circles are roughly in line with the filled circles, suggesting that this is a plausible hypothesis.

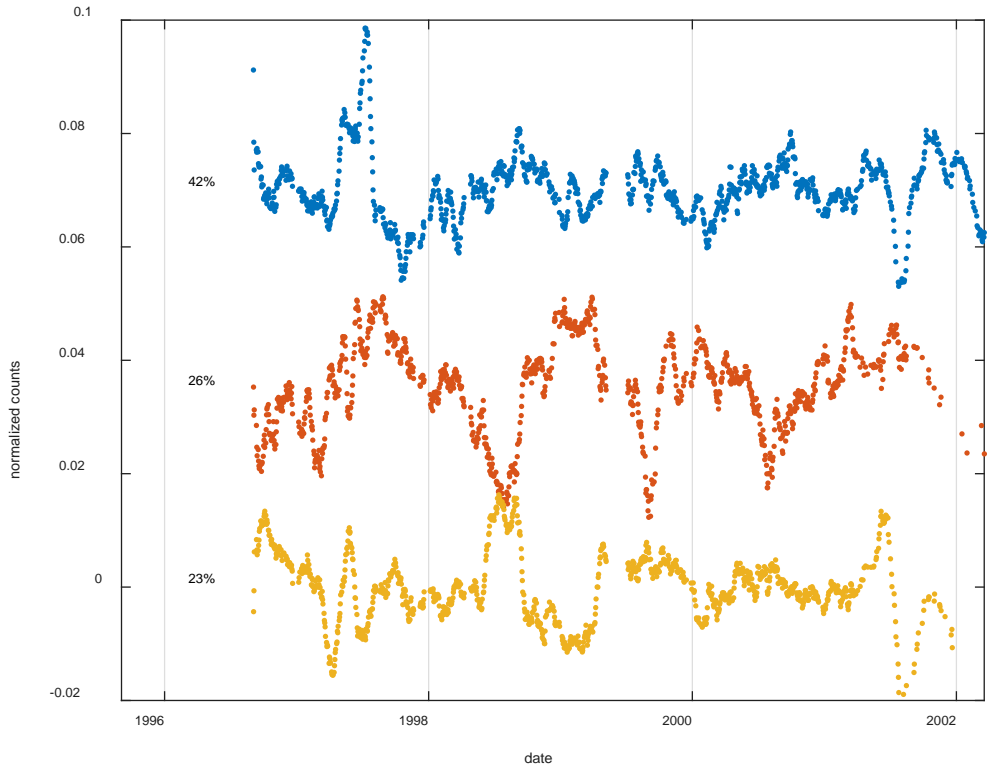

**Figure 16. Comparison of smoothed waveforms from the 23%, 26%, and 42% detectors.**

Sturrock et al [44] have given a range of phases that would be possible if a north-south asymmetry in the solar emission exists in addition to the variation in distance. Based on this hypothesis, phases in the range 0.183-0.683 (normalized to 1), i.e., March 8 to September 6, are forbidden. This "forbidden zone" corresponds to the region to the left of the diagonal dashed line in Figure 17. Note that the signals with inverted phase in Figure 19 are outside of this range. Thus if we accept the Sturrock range and the hypothesis that signals are present, these phases must be inverted by some instrumental or data processing step, since adding a phase of $\pi$ (182.62 days) brings them back into the permitted zone.

It is also interesting that the annual variation is only apparent over limited time intervals, with long intervals showing no significant component. If the variations are of solar origin, then this may be partially explained by variability in the phenomenon itself, as suggested earlier. However, this does not fully explain the presence or absence of variations in our data. The 21% data set is perhaps of too limited duration to make a definitive statement, but it is also puzzling why the data from only 4 of the remaining 7 detector systems exhibit a clear annual variation, when all of the detectors and associated systems were very similar with significant overlaps in data set dates.

Although it is likely that the measurements are influenced by other factors, it is clear that the overwhelming majority of events counted are the result of the activity of the sample. Consequently, since the annual variations do not appear in all data sets, we can conclude that the observed annual variations are not caused by variations of the activity. This is a much stronger conclusion than previously reported negative results, since the absence of variations over a specific relatively short time period could potentially be explained by periods of relative inactivity in solar processes responsible for the variation, or by differing sensitivities of isotopes to the proposed phenomenon.

At the same time, the annual variations almost certainly have a solar origin, though the mechanism is likely an environmental influence that has not yet been satisfactorily identified. We can therefore continue to consider the extent of solar influence, though the hypothesis of "new physics" is not required. Solar activity is known to have significant effects on the conditions in the upper atmosphere ("space weather"), and so the presence of more subtle effects near the earth's surface cannot be ruled out. However, Javorsek et al [39] compared the phases of data sets from PTB and BNL with the phases of local outdoor temperature, atmospheric pressure, relative humidity, earth-sun distance and their reciprocals, and concluded that none of these alone could explain the observed phases.

Phenomenologically, our results seem to suggest a system adjustment that can cancel the appearance of the annual variation. If the cancellation is not complete, then a small in-phase component will be present, and if there is over-compensation, the annual variation can appear with an inverted sign. For example, suppose the variation is present in the continuum pedestal of the gamma-ray spectrum rather than a specific energy line. This continuum is primarily the result of gamma rays that are not completely absorbed within the detector volume, and is subtracted from the total counts to estimate the net counts in the 609 keV line. The accuracy with which the amplitude of this pedestal is estimated would determine the amplitude and phase of any variation remaining after subtraction.

Cosmic ray muons are known to contribute to the continuum background, and the intensity at the earth's surface is known to exhibit a number of variations not unlike those that have been recently attributed to intrinsic variations of the activity. These include seasonal variations associated with atmospheric temperature [45],[46], and modulation of the cosmic ray intensity from solar coronal mass ejections resulting in decreases in cosmic ray intensity following solar flares (Forbush events) [47],[48],[49]; variations with the periodicity of the sidereal rotation rate [48],[45]; and variations correlated with the 11 year sunspot cycle [50],[48],[51]. Although these observations are highly suggestive, this contribution is expected to be orders of magnitude too small to explain the periodic variations observed (e.g., see [52], [20]). As an upper limit in our experiments, the continuum background within the region of interest for the 609 keV line without the sample present resulted in a rate equivalent to 7-20 counts in a 5 minute observation period. The largest component of this background is most likely radioactive contamination of the shield and parts of the spectrometer, but presumably contains a cosmic ray contribution as well. The amplitude of any residual contribution to the line area from this continuum background should have been further reduced substantially by the subtraction process, resulting in a contribution perhaps an order of magnitude smaller. Consequently it appears that contributions from cosmic rays to the PULSTAR measurements are between 1 and 2 orders of magnitude too small to explain the observed variations.

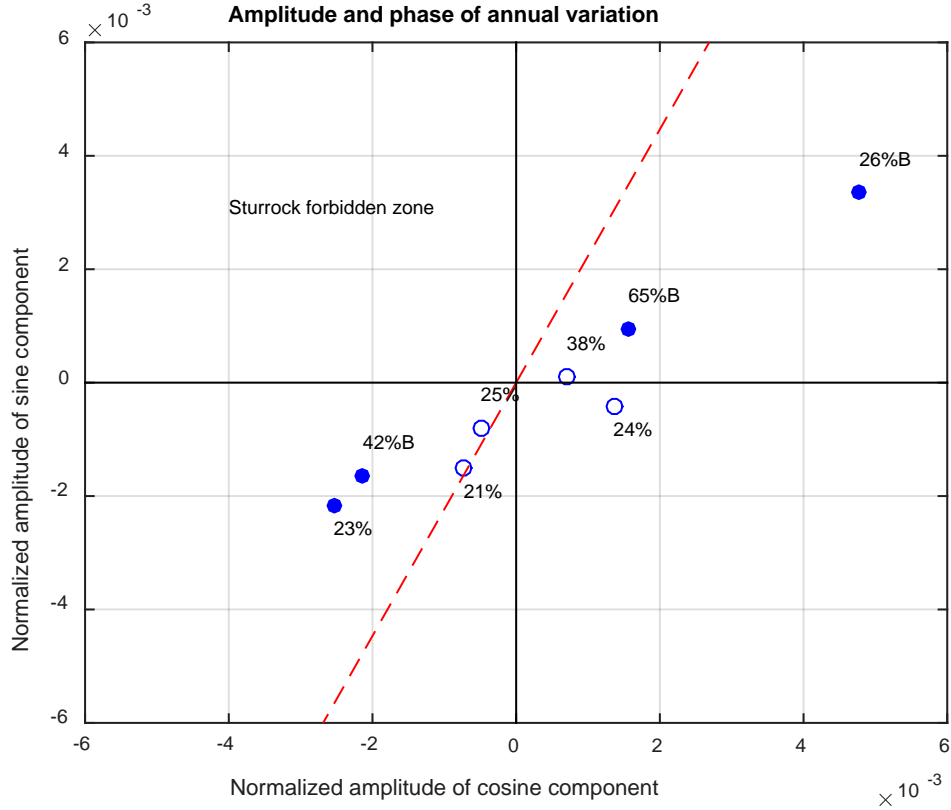

**Figure 17. Polar plot of periodogram components with a frequency of 1 cpy.**

It is also interesting that, in agreement with others [9-12], [17], [35], the short-time periodograms show that frequency components are not constant at 1 cpy, but can vary with time. To some extent, this apparent variation in frequency may simply be the result of frequency estimation errors in noise as described in Eqn. (2). However, in the context of the hypothesis that the variations are of solar origin, we would not expect the variations to have precisely a frequency of 1 cpy owing to dynamic processes in the sun [53]. For example, if the flux is modulated by dynamic processes in the sun, then the functional form of the flux would be approximately

$$S(t) \sim \frac{f(t)}{\left(R - \Delta R \cos(2\pi\Omega t - \phi)\right)^2}$$
$$\sim \frac{f(t)}{R^2}\left(1 + 2\frac{\Delta R}{R}\cos(2\pi\Omega t - \phi)\right), \quad (3)$$

where the cosine term describes the annual variation in distance between the earth and sun (with $\Omega \approx 1$ cpy), $R$ is the average earth-sun distance, and $f(t)$ describes the dependence of the flux on solar processes, as yet unspecified. Clearly the product term $f(t)\cos(2\pi\Omega t - \phi)$ will result in frequency components *other* than 1 cpy, depending on the variation of $f(t)$. As an example, if $f(t)$ varied on the scale of the 11-year sunspot cycle, then the observed frequency component could appear at about $1 \pm 0.09$ cpy. Combining these frequencies with the frequency uncertainty estimates from Table 5 yields values comparable to the observed frequency components nearest 1 cpy also given in Table 5. The time-varying behavior could also result in slippage of the phase of the variation with time, causing the observed variation in counts to go in and out of phase with the earth-sun distance, suggesting an additional mechanism for phase values outside of the Sturrock range.

Since there has also been discussion of the possibility of other frequency components in the periodograms [7], [12], [14], [15], [32], joint power statistics were constructed from the four data sets individually showing the presence of the annual variation. Since the 23%, 26%-B, and 42%-B data sets showed the presence of an annual variation over approximately the same time period, the joint power statistic (JPS) $J_3$ was constructed from these data sets using the approximation [54]

$$J_3 = \frac{2.916 X_3^2}{1.022 + X_3}, \tag{4}$$

Where $X_3$ is the geometric mean of the periodogram amplitudes

$$X_3 = \left( S_{23\%} S_{26\% B} S_{42\% B} \right)^{1/3} . \tag{5}$$

Similarly, the joint power statistic $J_4$ was constructed using all four data sets using [54]

$$J_4 = \frac{3.881 X_4^2}{1.269 + X_4}, \tag{6}$$

Where $X_4$ is the geometric mean

$$X_4 = \left( S_{23\%} S_{26\% B} S_{42\% B} S_{65\% B} \right)^{1/4} . \tag{7}$$

The joint power statistics are shown in Figure 18. The statistics $J_3$ and $J_4$ are very similar, exhibiting main peaks at 1.06 and 1.02 cpy, respectively, with secondary peaks at about 1.8 and 3 cpy. The secondary peaks are statistically significant, but we presently do not have a conjecture as to their origin.

Sturrock et al have examined several radioactive decay data sets for frequency components that could be associated with solar rotation. Their analysis of the BNL data taken in the 1982-85 time frame and the PTB taken roughly between 1984 and 1998 showed the presence of peaks at 11.25 cpy [7], [8], [32]. If interpreted as being associated with solar rotation as observed from earth (synodic rotation rate), the corresponding sidereal rotation rate would be 12.25 cpy. In subsequent analysis of measurements of the Sr-90 decay from Lomonov Moscow State University (LMSU) taken between 2002-2009 [14], periodogram peaks were identified that were associated with possible r-mode oscillations in the sun [55], [56]. If the r-mode oscillations interact with a structure such as a magnetic flux tube that is co-rotating with the sun, the frequencies as observed from the earth are given by [15]

$$\nu(l,m) = \frac{2m\nu_R}{l(l+1)} \tag{8}$$

with $m = 1$ and $l = 2, 3, \ldots$ The sidereal rotation frequency that best fit the LMSU data was $\nu_R = 12.08$ cpy. A second data set from PTB taken over the time period 1999-2008 also showed peaks that agreed with r-mode oscillations with $\nu_R = 12.08$ cpy.

To examine the possibility of these frequencies also being present in the PULSTAR data, we are interested in frequency components near 11 cpy in the periodograms for $J_3$ and $J_4$. From Figure 18 we see that there are no statistically significant peaks in the range 10-15 cpy. The statistic $J_3$ is of particular interest since it is derived from data sets covering approximately the same time period (roughly 1997-2002). From Figure 19(b) we see that the largest peak appears at 10.74 cpy, corresponding to a sidereal rotation rate of 11.74 cpy assuming a solar origin. The r-mode oscillation frequencies for $l = 2, 3 \ldots 6$ are indicated by arrows in Figure 19(a) for the JPS $J_3$, using the sidereal rotation rate $\nu_R = 11.74$ cpy. The frequencies for $l > 6$ would not be visible since the process used to remove the long term trend eliminated frequencies below $f_{low}$ (see Table 3). The periodogram in Figure 19 differs from that in Figure 18, however, in that a lower value of $f_{low} = 0.3$ cpy was used to minimize distortion of the periodogram near the frequency for $l = 6$. Statistically significant peaks are seen near the frequencies for $l = 5, 6$, and there is a peak near the frequency for $l = 3$ that is very near the PFA threshold of 0.05. No peaks are observed near the frequencies for $l = 2, 4$.

Frequencies in this range observed by others include a 13.53 cpy (period of 27 days) observed by Baurov et al [57], and 12.38 cpy (period of 29.5 days) observed by Baurov et al [58] in background measurements with a Ge(Li) detector in Troitsk. Neither of these frequencies appear in our Joint Power Statistics.

The implied sidereal rotation rate from the PULSTAR analysis of 11.74 cpy is close enough to the values of 12.08 for the LMSU data and second PTB data set, and 12.25 reported for the earlier PTB and BNL data to at least be suggestive. The value from the PULSTAR data is closest to that from the LMSU and later PTB data, and these data sets also have the largest temporal overlap. Consequently, these observations are consistent with association with solar rotational modes whose frequencies fluctuate slightly with time.

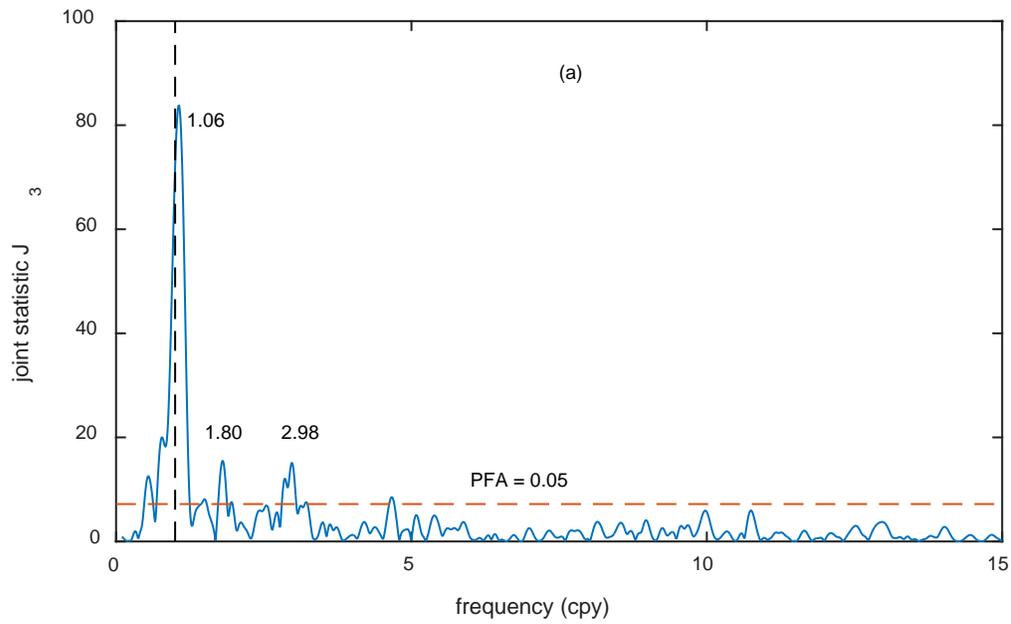

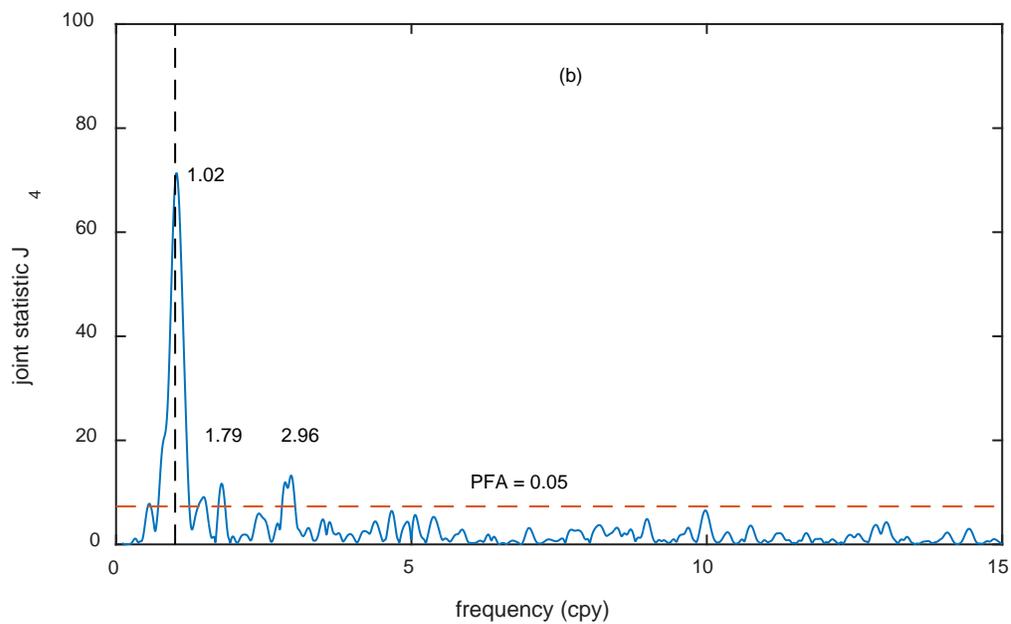

**Figure 18. The Joint Power Statistics for data sets exhibiting an annual variation. (a) The Joint Power Statistic from the 23%, 26%-B, and 42%-B data sets. (b) The Joint Power Statistic from the 23%, 26%-B, 42%-B, and 65%-B data sets. The vertical dashed line marks the frequency of 1 cpy.**

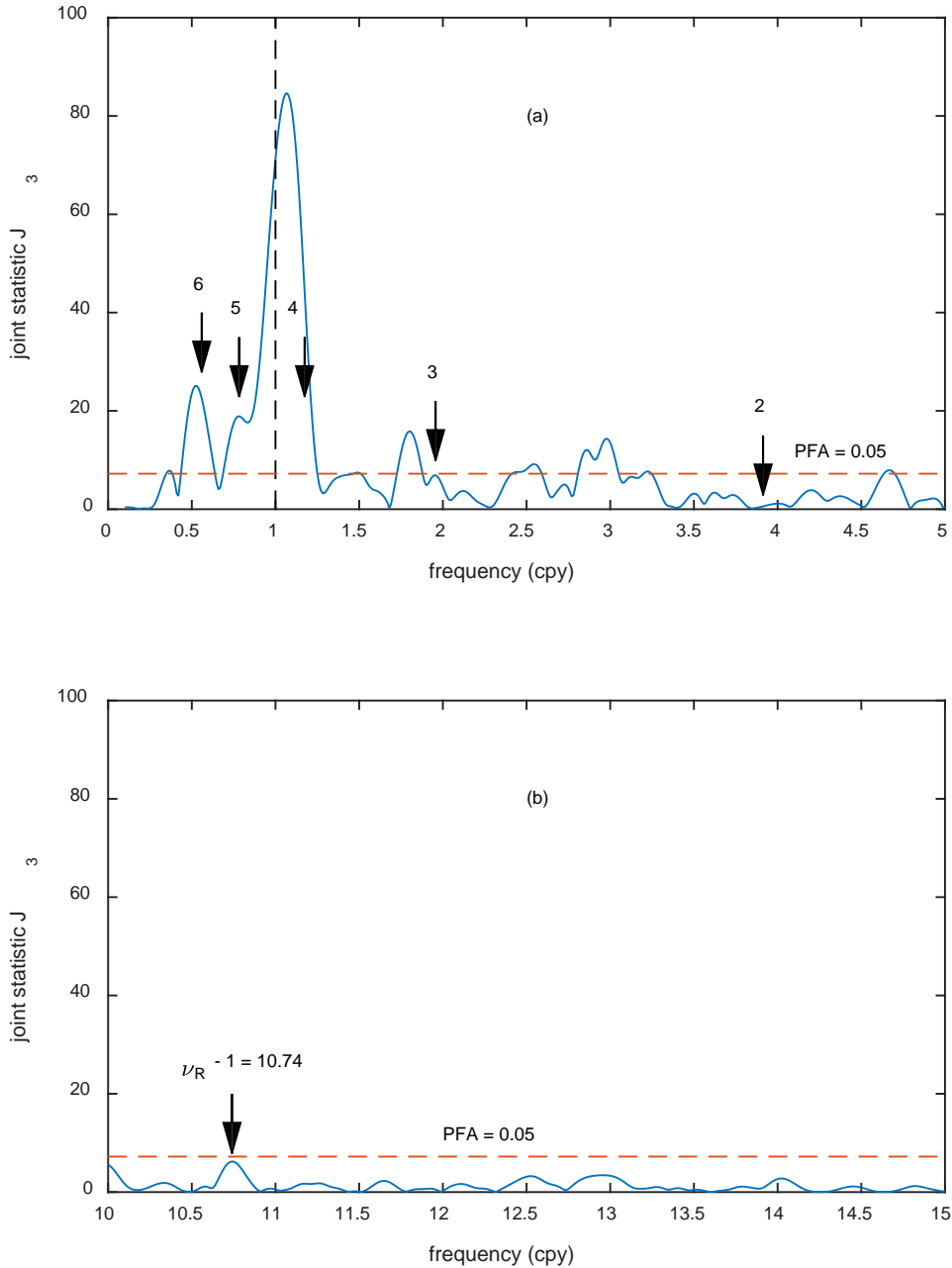

**Figure 19.** (a) r-mode oscillation frequencies for $\nu_R = 11.74$ cpy and $m = 1$ for $l = 2, 3 \ldots 6$. (b) Periodogram in the vicinity of the effective sidereal frequency as seen from the earth $\nu_R - 1 = 10.74$. These curves differ from those in Figure 18(a) in that the low-frequency cutoff was taken to be $f_{low} = 0.3$ instead of the value given in Table 3 ( $f_{low} = 0.5$ ). The low frequency analysis was extended to avoid truncating the peak for $n = 6$.

## 6. Conclusions

Eight data sets comprised of gamma ray counts at 609 keV from a standard Ra-226 sample as part of the quality control procedures at the NC State PULSTAR reactor were examined for the presence or absence of a variation with a period of 1 year. Statistically significant components were observed in four of the eight data sets. The phases of the signals are such that two have maxima about 30 days after 1 January, and the remaining two have maxima close to six months later. Three of the data sets have overlapping time intervals, so an explanation of a phase drift with time is ruled out. The amplitudes of the variation are in the range of 0.2%-0.5%. Both the actual frequencies and amplitudes of these signals were observed to vary with time, in agreement with other observations. Although we have not been able to determine how such a phase shift could arise from either the electronics or data processing, the observation of opposing phases by similar detectors over the same time period strongly suggests that this phase difference has an instrumental origin. More significantly, the fact that statistically significant annual variations were not observed in all data sets over similar time periods definitively eliminates variations in the activity as the cause of the variations. Hence there is no need to invoke new physics.

Environmental background measurements made with one of the detectors at a time near the peak of the annual variation resulted in counts that were too small by more than an order of magnitude to explain the observations. Consequently, an explanation in terms of seasonal variations in environmental radon does not appear plausible.

Solar activity is known to influence many phenomena on earth, and so even though we can conclude that there is no influence on the rate of radioactive decay, the annual variation in the measurement can be considered to be a solar influence. Since the sun is a complex dynamical system, it is reasonable to expect that the annual variation would be modulated by other solar processes, imposing additional temporal variations on the perturbing signal. Thus the possibility that other periodic components in the measurements are correlated with solar processes remains an interesting hypothesis. Our observations and analysis of these PULSTAR data sets are not inconsistent with this hypothesis.

Clear, quantitative explanations of the variations that have now been observed at multiple locations, and with various isotopes and detector types, are not yet available. Until the sources of these variations are understood, measurements of decay rates at a precision greater than 0.5% cannot be made with confidence. As pointed out by Ellis [2], in addition to the scientific significance, this uncertainty could have important implications for the interpretation of longitudinal medical studies.

A primary shortcoming of the incidental data sets that have been reported and studied is that simultaneous local environmental parameters were neither recorded nor controlled. To address this, intentional long term measurements (on the order of a decade) are needed in multiple locations with the same isotopes, similar detectors, and simultaneous recording of environmental parameters. At least two new experiments have begun with the purpose of making such long-term decay measurements, one in Utah in the United States [59], and one in the United Arab Emirates [60]. The Utah experiment is contained in an environmental chamber with temperature, pressure, and humidity kept constant, while the ambient temperature, pressure, and humidity are logged in the UAE experiment. Of particular interest are the common aspects of these experiments: measurements on Mn-54, Co-60, and Sr-90; and use of both Geiger-Müller and NaI detectors. However, neither long-term experiment employs a HPGe detector.

Finally, it may also be useful to perform a careful examination of the electronics and processing algorithms to search for unexpected mechanisms for susceptibility to environmental influences.

## Acknowledgments

We are indebted to Mr. Scott Lassell of the PULSTAR Reactor staff for access to the data records and assistance with its interpretation. C. R. G. acknowledges support by the US Department of Energy, Office of Nuclear Physics, under Grant No.~DE-FG02-97ER41041 (NC State University).

## References

[1] Alburger E, Harbottle G, Norton F, Half-life of 32Si, Earth and Planetary Science Letters 78 (2-3) (1986) 168-176, doi: 10.1016/0012-821X(86)90058-0.
[2] Ellis KJ, The effective half-life of a broad beam 238Pu/Be total body neutron irradiator, Physics in Medicine and Biology 35 (8) (1990) 1079-1088, doi: 10.1088/0031-9155/35/8/004.
[3] Siegert H, Schrader H, Schötzig U, Half-life measurements of Europium radionuclides and the long-term stability of detectors, Applied Radiation and Isotopes 49 (9-11) (1998) 1397-1401, doi: 10.1016/S0969-8043(97)10082-3.
[4] Falkenberg ED, Radioactive Decay Caused by Neutrinos? Apeiron 8 (2) (2001) 32-45.
[5] Fischbach E, Buncher B, Gruenwald T, Jenkins H, Krause E, Mattes J, Newport R, Time-Dependent Nuclear Decay Parameters: New Evidence for New Forces? Space Science Reviews 145 (3-4) (2009) 285-335, doi: 10.1007/s11214-009-9518-5.
[6] Jenkins JH, Fischbach E, Buncher JB, Gruenwald JT, Krause DE, Mattes JJ, Evidence of correlations between nuclear decay rates and Earth–Sun distance, Astroparticle Physics 32 (1) (2009) 42-46, doi: 10.1016/j.astropartphys.2009.05.004.
[7] Sturrock PA, Buncher JB, Fischbach E, Gruenwald JT, Javorsek D, Jenkins JH, Lee RH, Mattes JJ, Newport JR, Power Spectrum Analysis of Physikalisch-Technische Bundesanstalt Decay-Rate Data: Evidence for Solar Rotational Modulation, Sol Phys 267 (2) (2010) 251-265, doi: 10.1007/s11207-010-9659-4.
[8] Sturrock A, Buncher B, Fischbach E, Gruenwald T, Javorsek II D, Jenkins H, Lee H, Mattes J, Newport R, Power spectrum analysis of BNL decay rate data, Astroparticle Physics 34 (2) (2010) 121-127, doi: 10.1016/j.astropartphys.2010.06.004.
[9] Jenkins H, Fischbach E, Javorsek D, Lee H, Sturrock A, Concerning the time dependence of the decay rate of 137Cs, Applied Radiation and Isotopes 74 (2013) 50-55, doi: 10.1016/j.apradiso.2012.12.010.
[10] Sturrock PA, Fischbach E, Jenkins J, Erratum: "Analysis of Beta-Decay Rates for Ag108, Ba133, Eu152, Eu154, Kr85, Ra226, AND Sr90, Measured at the Physikalisch-Technische

Bundesanstalt from 1990 to 1996" (2014, ApJ, 794, 42), The Astrophysical Journal 796 (2) (2014) 149, doi: 10.1088/0004-637X/796/2/149.

[11] Sturrock PA, Fischbach E, Jenkins J, Analysis of Beta-Decay Rates for Ag108, Ba133, Eu152, Eu154, Kr85, Ra226, AND Sr90, Measured at the Physikalisch-Technische Bundesanstalt from 1990 to 1996, The Astrophysical Journal 794 (1) (2014) 42, doi: 10.1088/0004-637X/794/1/42.

[12] Sturrock A, Fischbach E, Javorsek D, Jenkins H, Lee H, Nistor J, Scargle D, Comparative study of beta-decay data for eight nuclides measured at the Physikalisch-Technische Bundesanstalt, Astroparticle Physics 59 (2014) 47-58, doi: 10.1016/j.astropartphys.2014.04.006.

[13] Parkhomov A.G., Researches of alpha and beta radioactivity at long-term observations, arXiv.org (2010) 1-6.

[14] Sturrock A, Parkhomov G, Fischbach E, Jenkins H, Power spectrum analysis of LMSU (Lomonosov Moscow State University) nuclear decay-rate data: Further indication of r-mode oscillations in an inner solar tachocline, Astroparticle Physics 35 (11) (2012) 755-758, doi: 10.1016/j.astropartphys.2012.03.002.

[15] Sturrock A, Bertello L, Fischbach E, Javorsek D, Jenkins H, Kosovichev A, Parkhomov G, An analysis of apparent r-mode oscillations in solar activity, the solar diameter, the solar neutrino flux, and nuclear decay rates, with implications concerning the Sun's internal structure and rotation, and neutrino processes, Astroparticle Physics 42 (2013) 62-69, doi: 10.1016/j.astropartphys.2012.11.011.

[16] Steinitz G, Piatibratova O, Kotlarsky P, Possible effect of solar tides on radon signals, Journal of Environmental Radioactivity 102 (8) (2011) 749-765, doi: 10.1016/j.jenvrad.2011.04.002.

[17] Sturrock A, Steinitz G, Fischbach E, Javorsek D, Jenkins H, Analysis of gamma radiation from a radon source: Indications of a solar influence, Astroparticle Physics 36 (1) (2012) 18-25, doi: 10.1016/j.astropartphys.2012.04.009.

[18] Jenkins JH, Herminghuysen KR, Blue TE, Fischbach E, Javorsek D, Kauffman AC, Mundy DW, Sturrock PA, Talnagi JW, Additional experimental evidence for a solar influence on nuclear decay rates, Astroparticle Physics 37 (2012) 81-88, doi: 10.1016/j.astropartphys.2012.07.008.

[19] Alexeyev EN, Gavrilyuk YM, Gangapshev AM, Kazalov VV, Kuzminov VV, Panasenko SI, Ratkevich SS, Sources of the systematic errors in measurements of 214Po decay half-life time variations at the Baksan deep underground experiments, Phys. Part. Nuclei 46 (2) (2015) 157-165, doi: 10.1134/s1063779615020021.

[20] Bellotti E, Broggini C, Di Carlo G, Laubenstein M, Menegazzo R, Pietroni M, Search for time modulations in the decay rate of 40K and 232Th, Astroparticle Physics 61 (2015) 82-87, doi: 10.1016/j.astropartphys.2014.05.006.

[21] Norman EB, Browne E, Shugart HA, Joshi TH, Firestone RB, Evidence against correlations between nuclear decay rates and Earth–Sun distance, Astroparticle Physics 31 (2) (2009) 135-137, doi: 10.1016/j.astropartphys.2008.12.004.

[22] Bellotti E, Broggini C, Di Carlo G, Laubenstein M, Menegazzo R, Search for time dependence of the 137Cs decay constant, Physics Letters B 710 (1) (2012) 114-117, doi: 10.1016/j.physletb.2012.02.083.

[23] Hardy JC, Goodwin JR, Iacob VE, Do radioactive half-lives vary with the Earth-to-Sun distance? Appl Radiat Isot 70 (9) (2012) 1931-3, doi: 10.1016/j.apradiso.2012.02.021.

[24] Bellotti E, Broggini C, Di Carlo G, Laubenstein M, Menegazzo R, Precise measurement of the 222Rn half-life: A probe to monitor the stability of radioactivity, Physics Letters B 743 (2015) 526-530, doi: 10.1016/j.physletb.2015.03.021.

[25] Kossert K, Nähle OJ, Disproof of solar influence on the decay rates of 90Sr/90Y, Astroparticle Physics 69 (2015) 18-23, doi: 10.1016/j.astropartphys.2015.03.003.

[26] O'Keefe D, Morreale L, Lee H, Buncher JB, Jenkins H, Fischbach E, Gruenwald T, Javorsek D, Sturrock A, Spectral content of 22Na/44Ti decay data: implications for a solar influence, Astrophysics and Space Science 344 (2) (2013) 297-303, doi: 10.1007/s10509-012-1336-7.

[27] Jenkins JH, Mundy DW, Fischbach E, Analysis of environmental influences in nuclear half-life measurements exhibiting time-dependent decay rates, Nuclear Instruments and Methods in Physics Research Section A: Accelerators, Spectrometers, Detectors and Associated Equipment 620 (2-3) (2010) 332-342, doi: 10.1016/j.nima.2010.03.129.

[28] de Meijer RJ, Blaauw M, Smit FD, No evidence for antineutrinos significantly influencing exponential β+ decay, Appl Radiat Isot 69 (2) (2011) 320-6, doi: 10.1016/j.apradiso.2010.08.002.

[29] van Rooy MW, An investigation of a possible effect of reactor antineutrinos on the decay rate of 22Na, PhD Dissertation, Faculty of Science at Stellenbosch University Department of Physics, University of Stellenbosch, (March, 2015).

[30] Jenkins JH, Fischbach E, Perturbation of nuclear decay rates during the solar flare of 2006 December 13, Astroparticle Physics 31 (6) (2009) 407-411, doi: 10.1016/j.astropartphys.2009.04.005.

[31] Mohsinally T, Fancher S, Czerny M, Fischbach E, Gruenwald JT, Heim J, Jenkins JH, Nistor J, O'Keefe D, Evidence for correlations between fluctuations in 54Mn decay rates and solar storms, Astroparticle Physics 75 (2016) 29-37, doi: 10.1016/j.astropartphys.2015.10.007.

[32] Sturrock PA, Fischbach E, Jenkins JH, Further Evidence Suggestive of a Solar Influence on Nuclear Decay Rates, Sol Phys 272 (1) (2011) 1-10, doi: 10.1007/s11207-011-9807-5.

[33] Bellotti E, Broggini C, Di Carlo G, Laubenstein M, Menegazzo R, Search for correlations between solar flares and decay rate of radioactive nuclei, Physics Letters B 720 (1-3) (2013) 116-119, doi: 10.1016/j.physletb.2013.02.002.

[34] Javorsek D, Brewer C, Buncher B, Fischbach E, Gruenwald T, Heim J, Hoft W, Horan J, Kerford L, Kohler M, Lau J, Longman A, Mattes J, Mohsinally T, Newport R, Petrelli A, Stewart A, Jenkins H, Lee H, Morreale B, Morris B, Mudry R, O'Keefe D, Terry B, Silver A, Sturrock A, Study of nuclear decays during a solar eclipse: Thule Greenland 2008, Astrophysics and Space Science 342 (1) (2012) 9-13, doi: 10.1007/s10509-012-1148-9.

[35] Sturrock PA, Fischbach E, Parkhomov A, Scargle JD, Steinitz G, Concerning the variability of beta-decay measurements, arXiv preprint arXiv:1510.05996 (2015) .

[36] Schrader H, Half-life measurements of long-lived radionuclides--new data analysis and systematic effects, Appl Radiat Isot 68 (7-8) (2010) 1583-90, doi: 10.1016/j.apradiso.2009.11.033.

[37] Lomb NR, Least-squares frequency analysis of unequally spaced data, Astrophysics and space science 39 (2) (1976) 447-462.

[38] Scargle JD, Studies in astronomical time series analysis. II - Statistical aspects of spectral analysis of unevenly spaced data, The Astrophysical Journal 263 (1982) 835-853, doi: 10.1086/160554.

[39] Javorsek D, Sturrock A, Lasenby N, Lasenby N, Buncher B, Fischbach E, Gruenwald T, Hoft W, Horan J, Jenkins H, Kerford L, Lee H, Longman A, Mattes J, Morreale L, Morris B, Mudry N, Newport R, O'Keefe D, Petrelli A, Silver A, Stewart A, Terry B, Power spectrum analyses of nuclear decay rates, Astroparticle Physics 34 (3) (2010) 173-178, doi: 10.1016/j.astropartphys.2010.06.011.

[40] Kovács G, Frequency shift in Fourier analysis, Astrophysics and Space Science 78 (1) 175-188, doi: 10.1007/BF00654032.

[41] Horne JH, Baliunas SL, A prescription for period analysis of unevenly sampled time series, The Astrophysical Journal 302 (1986) 757-763, doi: 10.1086/164037.

[42] Ferraz-Mello S, Estimation of Periods from Unequally Spaced Observations, The Astronomical Journal 86 (1981) 619-624, doi: 10.1086/112924.

[43] Steinitz G, Kotlarsky P, Piatibratova O, Anomalous non-isotropic temporal variation of gamma-radiation from radon (progeny) within air in confined conditions, Geophysical Journal International 193 (3) (2013) 1110-1118, doi: 10.1093/gji/ggt057.

[44] Sturrock A, Buncher B, Fischbach E, Javorsek ii D, Jenkins H, Mattes J, Concerning the Phases of the Annual Variations of Nuclear Decay Rates, The Astrophysical Journal 737 (2) (2011) 65, doi: 10.1088/0004-637X/737/2/65.

[45] Hess VF, New Results of Cosmic-Ray Research, Terr. Mag. 41 (1936) 345-350.

[46] Duperier A, The Seasonal Variations of Cosmic-Ray Intensity and Temperature of the Atmosphere, Proceedings of the Royal Society of London A: Mathematical, Physical and Engineering Sciences 177 (969) (1941) 204-216, doi: 10.1098/rspa.1941.0007.

[47] Forbush SE, On the effects in cosmic-ray intensity observed during the recent magnetic storm, Physical Review 51 (12) (1937) 1108.

[48] Morrison P, Solar origin of cosmic-ray time variations, Physical Review 101 (4) (1956) 1397.

[49] Navia CE, Augusto CRA, Tsui KH, Robba MB, Mini-Forbush events on the muon flux at sea level, Physical Review D 72 (10) (2005) , doi: 10.1103/physrevd.72.103001.

[50] Forbush SE, World-wide cosmic ray variations, 1937--1952, Journal of Geophysical Research 59 (4) (1954) 525-542.

[51] Ahluwalia HS, Sunspot numbers, interplanetary magnetic field, and cosmic ray intensity at earth: Nexus for the twentieth century, Advances in Space Research 52 (12) (2013) 2112-2118, doi: 10.1016/j.asr.2013.09.009.

[52] Vojtyla P, Beer J, Šťavina P, Experimental and simulated cosmic muon induced background of a Ge spectrometer equipped with a top side anticoincidence proportional chamber, Nuclear Instruments and Methods in Physics Research Section B: Beam Interactions with Materials and Atoms 86 (3) (1994) 380-386.

[53] Fischbach E, Jenkins JH, Sturrock PA, Evidence for Time-Varying Nuclear Decay Rates: Experimental Results and Their Implications for New Physics, eprint arXiv:1106.1470 (2011) .

[54] Sturrock PA, Scargle JD, Walther G, Wheatland MS, Combined and comparative analysis of power spectra, Solar Physics 227 (1) (2005) 137-153.

[55] Papaloizou J, Pringle JE, Non-radial oscillations of rotating stars and their relevance to the short-period oscillations of cataclysmic variables, Monthly Notices of the Royal Astronomical Society 182 (3) (1978) 423-442, doi: 10.1093/mnras/182.3.423.

[56] Wolff CL, Blizard JB, Properties of r-modes in the Sun, Solar Physics 105 (1) 1-15, doi: 10.1007/BF00156371.

[57] Baurov YA, Sobolev YG, Ryabov YV, Kushniruk VF, Experimental investigations of changes in the rate of beta decay of radioactive elements, Physics of Atomic Nuclei 70 (11) (2007) 1825-1835, doi: 10.1134/s1063778807110014.

[58] Baurov YA, Konradov AA, Kushniruk VF, Kuznetsov EA, Sobolev YG, Ryabov YV, Senkevich AP, Zadorozsny SV, Experimental Investigations of Changes in β-Decay Rate of 60 Co and 137 Cs, Modern Physics Letters A 16 (32) (2001) 2089-2101.

[59] Ware MJ, Bergeson SD, Ellsworth JE, Groesbeck M, Hansen JE, Pace D, Peatross J, Instrument for precision long-term β-decay rate measurements, Rev Sci Instrum 86 (7) (2015) 073505, doi: 10.1063/1.4926346.

[60] Goddard B, Hitt GW, Solodov AA, Bridi D, Isakovic AF, El-Khazali R, Abulail A, Experimental setup and commissioning baseline study in search of time-variations in beta-decay half-lives, Nuclear Instruments and Methods in Physics Research Section A: Accelerators, Spectrometers, Detectors and Associated Equipment 812 (2016) 60-67, doi: 10.1016/j.nima.2015.12.026.